# Inferring Exoplanet Disequilibria with Multivariate Information in Atmospheric Reaction Networks

Theresa Fisher[1†], Hyunju Kim[1,2,3†], Camerian Millsaps[1], Michael Line[1], Sara Walker[1,2,3,4*]

[1]School of Earth and Space Exploration, Arizona State University, Tempe AZ USA
[2]Beyond Center for Fundamental Concepts in Science, Arizona State University, Tempe AZ USA
[3]ASU-SFI Center for Biosocial Complex Systems, Arizona State University, Tempe AZ USA
[4]Santa Fe Institute, Santa Fe, NM USA

[†]these authors contributed equally to this work
[*]author for correspondence: sara.i.walker@asu.edu

# Abstract

**Inferring the properties of exoplanets from their atmospheres, while confronting low resolution and low signal-to-noise, poses rigorous demands upon the data collected from observation. Further compounding this challenge is that inferences of exoplanet properties are built from forward models, which can include errors due to incomplete or inaccurate assumptions in atmospheric physics and chemistry. The confluence of observational noise and model error makes developing techniques to identify predictive features robust to both low s/n and model error increasingly important for exoplanet science. We demonstrate how both issues can be addressed simultaneously by taking advantage of underutilized multivariate information present in atmospheric models, including thermodynamic statistics and reaction network structure. We provide a case study of the prediction of vertical mixing (parameterized as eddy diffusion) in hot Jupiter atmospheres and show how prediction efficacy depends on what model information is used - e.g. chemical species abundances, network statistics, and/or thermodynamic statistics. We also show how the most predictive variables vary with planetary properties such as temperature and metallicity. Our results demonstrate how inferences built on single metrics do not have utility across all possible use cases. We discuss future directions applying multivariate and network model information as a framework for increasing confidence in inferences aimed at extracting features relevant to exoplanet atmospheres, and future applications to the detection of life on terrestrial worlds.**

# Introduction

Exoplanets exhibit a diverse range of physical properties (temperature, mass, composition, etc.) far exceeding that observed within our own solar system (Seager 2013). Characterizing these unknown worlds based on limited and low resolution data presents a significant challenge, but the grander challenge will be to eventually identify and characterize the unknown life that might inhabit some of them (Walker et al. 2017). The lack of solar system analogs for the majority of exoplanets discovered means we do not have nearby analogs for most worlds discovered, and that our knowledge about these will be based entirely on the efficacy of our tools for remote inference. The comparatively low resolution and signal to noise data available from current and near-term technology mean we may have to develop new methods to meet these challenges (Fujii et al. 2018).

Current state-of-the-art inference approaches include Bayesian retrieval methods, that extract atmospheric information from spectra by first forward-modeling atmospheric chemistry under reasonable model assumptions and then fitting the predictions of those forward models to data through a parameterized model of the atmosphere combined with parameter estimation tools like Markov chain Monte Carlo's (Benneke & Seager 2012; Line et al. 2015; Madhusudhan 2019; Madhusudhan et al. 2011). A limitation of this approach is that our ability to infer properties of exoplanets is heavily dependent on the accuracy of the assumptions in the forward-models in capturing the relevant planetary physics and chemistry - a steep challenge given the aforementioned lack of solar system analogs for most exoplanets. Recently, there has also been intensifying interest in the application of machine learning approaches to retrieval methods (Cobb et al. 2019; Hayes et al. 2020). These also allow making predictive inferences if trained on reliable model data; however, like Bayesian retrieval methods, this requires first having accurate models to generate the training data sets.

The combination of these factors suggests that developing techniques to identify predictive features in atmospheric models, or developing new frameworks entirely, which are robust to both low s/n and to model error will become increasingly important as we search for atmospheric biosignatures beyond our solar system in the decades to come (Catling et al. 2018; Kiang et al. 2018; Walker et al. 2017). We demonstrate how many of these issues can start to be addressed by taking advantage of the multivariate information present in atmospheric models including chemical composition, thermodynamic statistics, and atmospheric reaction network topology, the latter of which is agnostic to specific details of atmospheric chemistry. We test these frameworks by exploring the chemical disequilibria of hot Jupiters in terms of a relatively well-understood driver of chemical disequilibrium, the vertical mixing coefficient, $K_{zz}$. We show how some variable statistics or combinations thereof are better predictors of $K_{zz}$ in hot Jupiter atmospheres than others, and might therefore be the best targets for the development of new inference methods.

In what follows, we apply this approach to hot Jupiters. Our motivation in this choice is threefold: (1) these remain one of the few classes of exoplanets with relatively well-characterized atmospheres (Crossfield 2015; Madhusudhan 2019; Seager & Deming 2010); (2) many of the chemical kinetics involved have been thoroughly investigated in laboratory settings (Fortney et al. 2021; Venot et al. 2012); and (3) a relatively large number of example planets are known. Disequilibria are a prime target for developing new frameworks for remote inference because they allow characterizing planetary conditions, and because it is speculated they can be driven by life (for terrestrial worlds) (Krissansen-Totton et al. 2016). These features motivate our focus in this study on inference of $K_{zz}$ as a model for how multivariate approaches can provide more reliable data for inference. In Section 1, we show that by tracking statistics of interactions among species in an atmosphere (e.g. via its network representation), we can better constrain the likely $K_{zz}$ value with much greater resolution than with just the observable species data alone. In Section 2, we also show how network properties tend to be more robust to observational uncertainty or perturbations representing data missing from the model. In Section 3, we discuss implications for the development of new inference methods that directly target the most informative atmospheric variables. Overall our results indicate that a multivariate analysis, treating atmospheres as complex systems and leveraging quantitative frameworks from complexity science, allows new windows into understanding exoplanet atmospheric properties from remote data.

## Analyzing Multivariate Information from Atmospheric Models

Forward modeling approaches use thermochemical reactions and their rates to simulate possible exoplanet atmospheres with specified temperature, composition, and metallicity. Using a grid of models, likelihood distributions of key species abundances can be predicted and used for inferring planetary properties, such as vertical mixing strength, $K_{zz}$, from atmospheric data, see Figure 1 (left panel). However, other features can be derived from the same models that could be used to build more accurate inferences from limited data. This includes properties of the statistics of interactions among molecules within an atmospheric chemical reaction network, as outlined in Figure 1 (right panel), which can be captured quantitatively by representing the chemical reactions in a planetary atmosphere as a complex network (Centler & Dittrich 2007; Gleiss et al. 2001; Newman et al. 2011). It also includes statistics of the thermodynamics of reactions, which are related directly to measures of atmospheric disequilibria (Krissansen-Totton et al. 2018); (Krissansen-Totton et al. 2016; Line & Yung 2013; Molaverdikhani et al. 2019; Simoncini et al. 2013). Network theoretic and thermodynamic measures have been proposed as possible candidate measures of global biological activity on terrestrial worlds (Krissansen-Totton et al. 2016; Walker et al. 2017) and could provide new targets for direct inference. In what follows, we adopt a multivariate approach leveraging information in the thermodynamic statistics and the network topology of planetary atmospheres, generated from simulated atmospheric data. Our goal is to determine what information about the disequilibria associated with vertical mixing is captured in different statistics characterizing atmospheric properties in terms of species abundances, thermodynamics, and network structure.

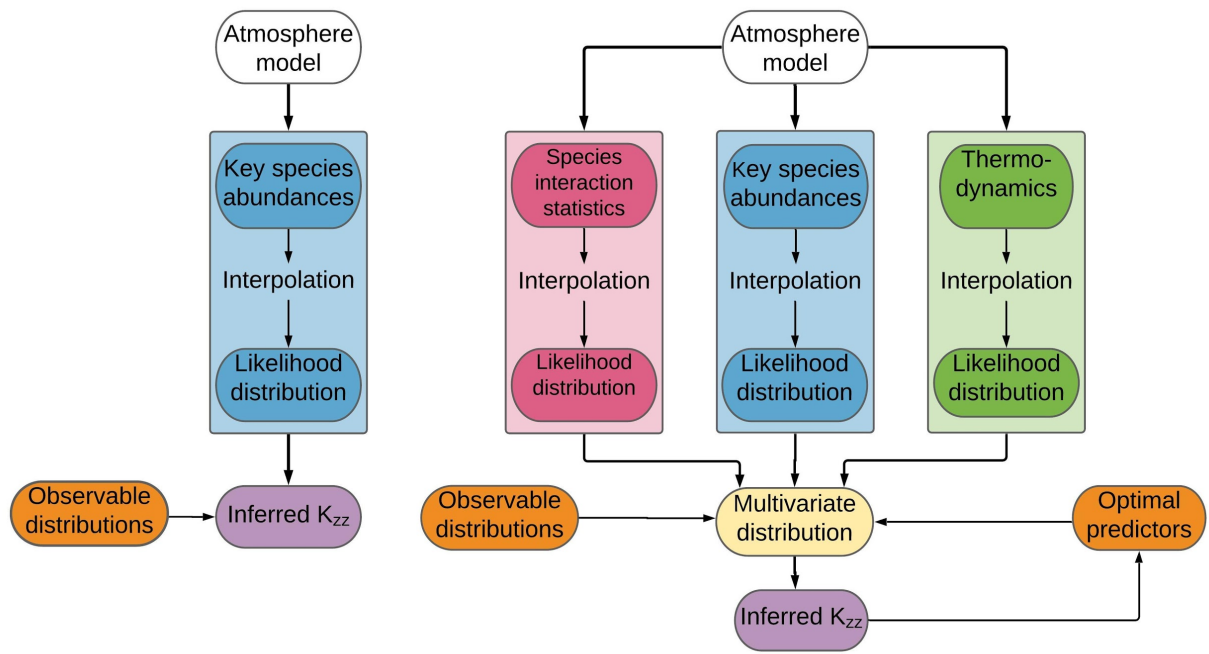

*Figure 1:* Left: "Standard" analysis pipeline used to infer atmospheric $K_{zz}$ by generating a likelihood distribution from the abundance of key species. Right: Schematic of the analysis pipeline used for the multivariate approach. Inverse modeling, based on observational data, is used to create a model of a planet's atmospheric chemistry. From this model, the list of thermodynamically favorable reactions and their rates is used to construct the chemical reaction network, which is then analyzed and its topological, thermodynamic, and species abundance parameters measured. The parameters then form the basis of an interpolation function, which is fed a selection of initial physical conditions, drawn from a Gaussian distribution, to determine what the corresponding network parameters are if a network was to be constructed using those initial physical conditions. The resulting distributions can then be used to compare any given atmosphere with another in terms of distance from thermochemical disequilibrium and can aid future inverse modeling.

## Atmospheric Modeling with VULCAN

We used the VULCAN modeling package (Tsai et al. 2017) to simulate the atmospheric chemistry of hot Jupiters over a grid of temperatures (400K to 3000K, in ~100K increments), metallicities (incremented as 0.1, 1, 3, 10, 30, 100, and 300 times that of the solar system, evaluated on a log scale), pressures (from 50mb to 150mb, reflecting the likely depth in the atmosphere that will be observable, in increments of ~50 mb), and finally, vertical mixing coefficients, with values of $K_{zz}$

= 0, $10^6$, $10^8$, and $10^{10}$ s$^{-1}$ cm$^{-1}$. In the rest of the paper, the unit of $K_{zz}$ is assumed to be s$^{-1}$ cm$^{-2}$ unless it is mentioned otherwise. This made for a total of 1,882 model runs simulated over the grid.

The vertical mixing coefficient ($K_{zz}$) represents how much mixing is occurring between the lower layers of the atmosphere, which are at higher temperature and pressure, and the upper layers, which are at lower temperature and pressure. At a $K_{zz}=0$, there is no mixing between layers, and each layer is assumed to very quickly reach thermochemical equilibrium; thus, we consider this to be our equilibrium scenario. As the atmosphere becomes more mixed, it moves away from equilibrium because molecules are produced at the lower levels of the atmosphere in quantities that are not thermodynamically favorable at the higher levels, and then "dredged" up to the higher levels at a rate faster than they can be depleted by equilibrium chemistry. In other words, the rate of diffusion is higher than the rate of thermochemical equilibration. Thus, $K_{zz}$ is often used as a proxy of how far away the atmospheric chemistry is from equilibrium. The result of these models is thermochemical steady-state values for the abundances of all species included in the model over a range of pressures, see Figure 2 (left). While such measurements are familiar to planetary and atmospheric scientists, some of the multivariate features we track, such as properties describing the arrangement of connections between points in the atmospheric chemical reaction network shown in Figure 2 (right), provide more accurate inference of $K_{zz}$ than the abundances do. Such network measures might include *mean degree* (the average number of connections per point in a network) or *average shortest path length* (the average shortest distance between any two points in a network) (see Table 1 and the section, "Quantifying the Characteristic Topology of Chemical Reaction Networks"). As we will show in more detail, these other ways of quantifying atmospheric interactions can offer additional information over the same range of physical values.

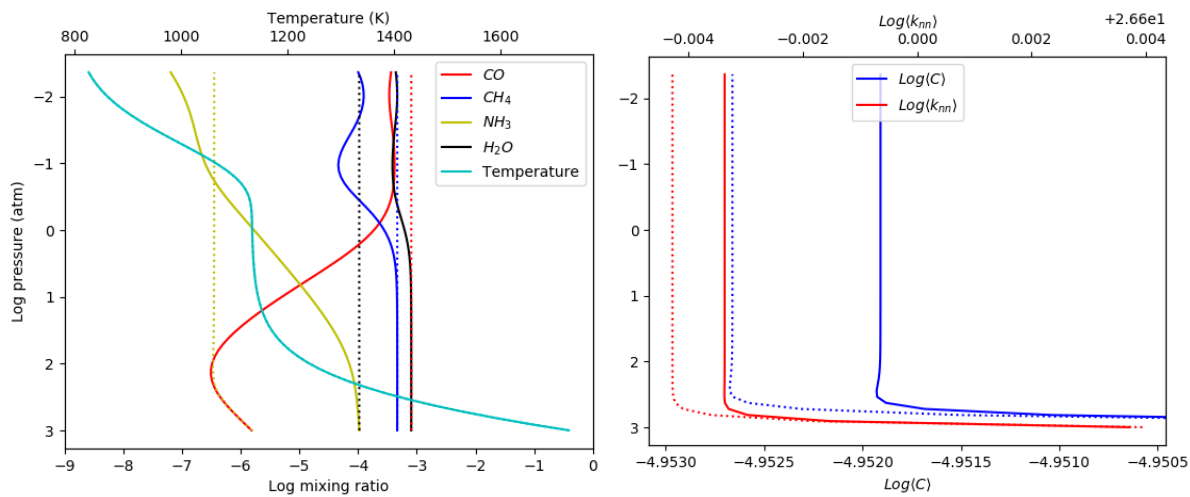

*Figure 2: Left: the abundance of key molecular species as a function of pressure, at 1200K and 1 solar metallicity. Right: Network topological measures - average clustering coefficient, and average neighbor degree (see Table 1 and section "Quantifying the Characteristic Topology of Chemical Reaction Networks") - as a function of pressure, at temperature 1200K and 1 solar metallicity. Network topological*

*measures vary as physical conditions, such as pressure, change. Solid lines represent the atmosphere at $K_{zz}$ of zero, dashed lines an atmosphere at $K_{zz}$ of $10^{10}$.*

## Quantifying the Characteristic Topology of Chemical Reaction Networks

We constructed chemical reaction networks (CRNs) from the simulated atmospheric data of hot Jupiters for each model on our grid. Systems of interacting components can be projected onto network representations to study statistical relationships in the interaction of these components. While network analyses are widespread in their application in the study of complex systems, such as biological or technological examples, they have so far seen less application to astronomical examples although a few notable examples do exist. Solé and Munteanu (Solé & Munteanu 2004) compared the topology of chemical reaction networks of the atmosphere of Earth to those of Titan, Mars, and Jupiter, and found that Earth's exhibited unique hierarchical and modular properties. The CRNs of the interstellar medium have also been analyzed (Jolley & Douglas 2010), and are distinguishable from the CRNs of biological systems (Jolley & Douglas 2012). Estrada (Estrada 2012) further evaluated network topology as a potential biosignature, and also introduced criteria of quantifying a network's ability to return to thermodynamic equilibrium.

Otherwise, however, most atmospheric modeling has focused strictly on recovering species abundances (e.g. the pipeline in the left panel of Figure 1). We introduce a network theoretic approach here to explore statistical analysis for the collective interdependent interactions among chemical species in exoplanetary atmospheres as an additional layer of information available to remote inference. We expect this will provide a pathway toward the development of new tools for atmospheric reaction network inference in exoplanet atmospheres of utility across different planet types. We are particularly motivated to develop these approaches for their longer-term applicability to terrestrial world atmospheres (Walker et al. 2017). In recent years it has become increasingly recognized that current biosignature gases such as $CH_4$ and $O_2$, can only be interpreted as biosignatures when considering their planetary context (Meadows 2017; Schwieterman et al. 2018). Inferring the global structure of a network from atmospheric data affords the possibility of not just tracking individual species abundances, but also systems-level patterns in chemical reactions happening in planetary atmospheres. Thus, understanding the properties of network representations of atmospheric chemistry has potential to provide the needed contextual measurements that directly probe how life can drive features of atmospheric chemistry at a systems-level (Walker et al. 2017). However, to even approach developing this as a method for biosignature detection we first need to investigate how a network-theoretic approach can contribute to understanding planetary atmospheres in the absence of life, a key motivation for our adopting this mathematical framework as part of our multivariate pipeline for studying hot Jupiter atmospheres.

We constructed network representations, also called graphs, of the chemical reactions for hot Jupiter atmosphere models, by first assigning each species present in the model as a node and linking two nodes by an edge if they co-participate in a reaction. See Figure 3 for a visualization of network projections at different temperatures and $K_{zz}$. Edges in our networks are weighted by the rate of the reaction the edge represents. Rates were calculated by taking the product of the concentration of reactants and the reaction rate constant at a steady state for the reaction defining the edge (note because more than two species typically participate in any given reaction, a single reaction can correspond to more than one edge in this network projection, where each edge associated to that reaction carries the same weight). Projecting real systems into abstract, graphical representations as complex networks is non-trivial and many choices are possible. For example, one could choose a bipartite representation where both chemical species and reactions are treated as two distinct classes of nodes (see e.g. (Montañez et al. 2010) for discussion in the context of biochemical networks of motivations behind different graph representations of the same system). We choose here to use a unipartite representation, where only chemical species are represented as nodes. It allows us to study interaction patterns among species within a chemical atmosphere and focus on identifying its statistical characteristics from a relatively simple network projection. The concentrations, reaction list, and reaction rate constants used for network construction were all taken directly from the VULCAN model specifications and model outputs at steady-state.

For each network generated on our model grid, to capture its global topological properties, we calculated the average of several standard network theoretic measures over a whole network: degree, shortest path length, clustering coefficient, average neighbor degree, node betweenness centrality, and edge betweenness centrality. The average value of these measures can quantify global structure emerging from collective interactions among chemical species in the atmosphere. We provide a summary of each measure in what follows. More detailed descriptions of these measures can be found in (Albert & Barabási 2002; Newman 2003, 2010) for general cases and (Antoniou & Tsompa 2008; Barrat et al. 2004; Onnela et al. 2005) with a particular focus on weighted networks.

Let $V$ and $E$ a set of nodes and edges for the atmospheric networks, respectively. The nodes represent chemical species participating in chemical reactions in the atmosphere and two nodes are connected in the network when two species participate in the same chemical reaction as a reactant and a product. It is important to note that the edges in our network representations are 'weighted'--that is, a value is assigned to each connection between two nodes. This weight of an edge connecting two nodes $i$ and $j$, $w_{ij}$ is determined by taking the reaction rate constant of the reaction the edge represents and multiplying it by the abundance of the reactants involved in the reaction.

***Degree*** is one of the most frequently used network measures to characterize the structure of a given network (Jeong et al. 2000; Smith et al. 2021). For an unweighted network, the degree of node $i$, $d_j$ is simply the number of edges connecting the nodes to the rest of the network. However, for a weighted network like the atmospheric network model in this paper, the degree of node $i$, $k_i$ is the sum of weights of the edges connected to $i$:

$$k_i = \sum_{j \in V} a_{ij} w_{ij}$$

where $a_{ij}$ is an element of an adjacency matrix of a given network and equal to 1 when $i$ and $j$ are connected by an edge and 0 otherwise. The degree quantifies the number of chemical species that a given species shares a reaction with (is connected by an edge with), with each weighted by the rate of their shared reaction. Roughly speaking, this captures how much of the flux of chemical species happens due to reactions that the given chemical species participates in. For our further analysis, we utilized the ***mean degree***, $\langle k \rangle$, defined as the average weight over all edges in a network:

$$\langle k \rangle = \frac{1}{N} \sum_{i \in V} k_i$$

where $N$ is the total number of nodes in the network. Mean degree represents the flux of chemical species averaged across all species in a given atmospheric network model.

***Shortest path length*** between two nodes $i$ and $j$ for a weighted network is defined as

$$\ell(i,j) = \min_{\gamma(i,j) \in \Gamma(i,j)} \left[ \sum_{u,v \in \gamma(i,j)} w_{uv} \right]$$

where $\gamma(i,j)$ is a path from $i$ to $j$ and $\Gamma(i,j)$ is the set of all possible paths from $i$ to $j$ (Antoniou & Tsompa 2008). In chemical reaction networks, it represents the smallest amount of flux along a pathway connecting any two species (note in a bipartite network the path would correspond to the more familiar definition of pathway as a linear sequence of reactions. However, in the unipartite representation, we use here, the path should instead be thought of co-occurrence or dependence of chemical species over reactions where the end member species $i$ and $j$ are participants in at least one reaction each). The **average shortest path length** over all pairs of nodes in the network can be written as

$$\langle \ell \rangle = \frac{1}{N(N-1)} \sum_{i,j \in V} \ell(i,j)$$

and this quantifies the average minimum weighted distance between any nodes.

***Clustering coefficient*** indicates the weighted density of edges between neighbors of a given node:

$$C_u = \frac{1}{d_u(d_u - 1)} \sum_{vw} (\hat{w}_{uv} \hat{w}_{uw} \hat{w}_{vw})^{1/3}$$

where $d_u$ is the number of edges connected to a node $u$, and $\hat{w}_{uv} = w_{uv}/max(w)$ is normalized by the maximum weight in the network (Onnela et al. 2005). $C_u$ is assigned to 0 for $d_u < 2$. A group of nodes with high clustering coefficient will be tightly knit. **Average clustering coefficient**,

$$\langle C \rangle = \frac{1}{N} \sum_{i \in V} C_i$$

is the average of clustering coefficient over all nodes and quantifies how densely the nodes' neighbors are interconnected. (Barabasi and Pófai 2016).

*Average neighbor degree* is the average weighted degree of the nodes connected to a given node (Barrat et al. 2004) and can be written as

$$k_{nn,i} = \frac{1}{k_i} \sum_{j \in N(i)} w_{ij} d_j$$

where $k_i$ is the degree of node $i$, $N(i)$ is the set of nodes connected to $i$, $w_{ij}$ is the weight of the edge that links nodes $i$ and $j$, and $d_j$ is the number of edges connected to node $j$. Note that $d_j$ is equivalent to $k_i$ in unweighted networks. The correlation between average neighbor degree and degree is useful for determining if a given network is assortative - high degree nodes tend to be connected to high degree ones and avoiding low degree ones - or disassortative - high degree nodes tend to be connected to low degree nodes. In atmospheric chemistry, it can identify statistically whether chemical species associated with a high total flux through them tend to participate in the same reactions. In this paper, we computed the **mean value of average neighbor degree** $\langle k_{nn} \rangle$ over all nodes in a given network for the statistical analysis. Note that in general the mean value of the average neighbor degree alone can not determine assortativity. However, in our atmospheric network models where a few nodes have a degree higher by several orders of magnitude than other nodes and dominate the total flux within the network, the high mean value can indicate the high degree nodes tend to be connected to each other.

*Node betweenness centrality* is defined as

$$g(v) = \sum_{s,t \in V} \frac{\sigma(s,t \mid v)}{\sigma(s,t)}$$

where $V$ is the set of nodes, $\sigma(s,t)$ is the number of shortest $(s,t)$-paths, and $\sigma(s,t|v)$ is the number of those paths passing through some node $v$ other than $s$, $t$. If $s = t$, $\sigma(s,t) = 1$, and if $v \in s,t$, $\sigma(s,t|v)=0$. In a chemical reaction network, this measure quantifies how important a given molecule is across multi-step reactions occurring within the network. It is important to note that nodes with high betweenness centrality can sometimes be low degree, but essential to dynamics and function since they play a key structural role by connecting many otherwise disconnected or distant nodes. $\langle g(v) \rangle$ represents the **average of *g(v)*** over all nodes within a given network.

*Edge betweenness centrality* represents the number of shortest paths going through that transverse through edge *e* and, similar to node betweenness centrality, is defined as

$$g(e) = \sum_{s,t \in V} \frac{\sigma(s,t \mid e)}{\sigma(s,t)}$$

where $V$ is the set of nodes, $\sigma(s,t)$ is the number of shortest $(s,t)$-paths, and $\sigma(s,t|e)$ is the number of those paths passing through edge *e*. $\langle g(e) \rangle$ represents the **average of $g(e)$** over all edges.

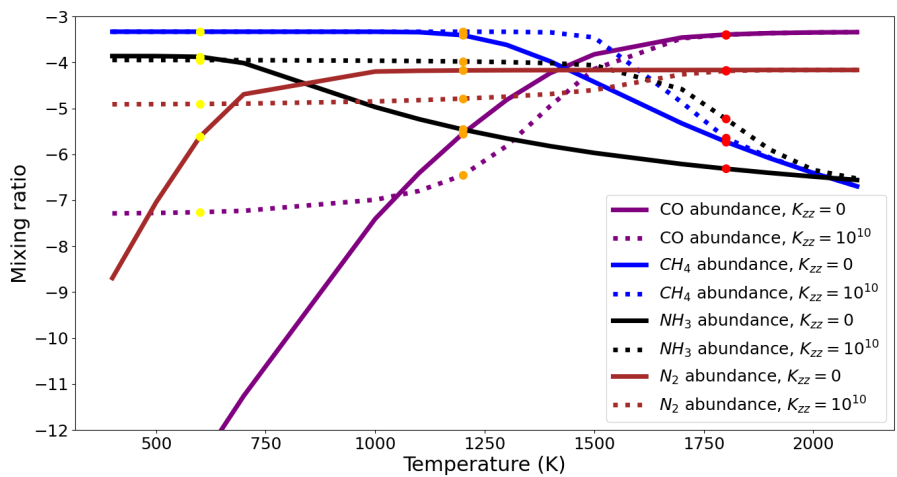

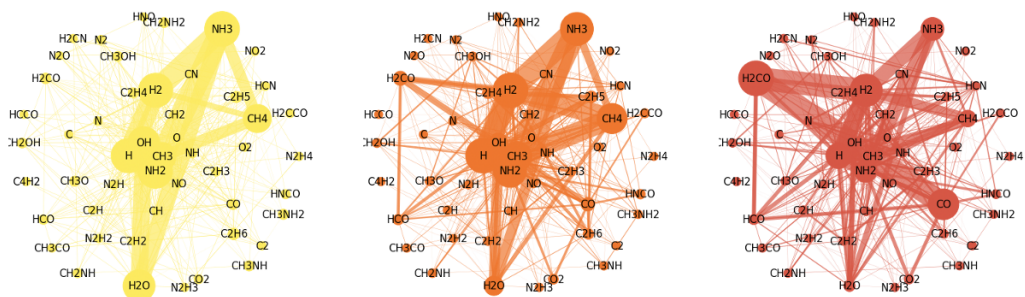

*Figure 3 :* Top: The abundance of $CH_4$, CO, $NH_3$, and $N_2$ as a function of temperature, demonstrating the transition from abundant reduced species ($CH_4$ and $NH_3$) to abundant oxidized species (CO and $N_2$) as temperatures increase. Increasing the $K_{zz}$ from equilibrium (solid) to $K_{zz}= 10^{10}$ (dashed lines) leads to the transition occurring at higher temperatures than the equilibrium case. The yellow, orange and red dots mark the temperatures of the corresponding networks shown in Figure 3b. Bottom: Network diagrams of the atmospheric chemical reaction network of a hot Jupiter over a range of temperatures, at solar metallicity and $K_{zz} = 0$ (top) and $K_{zz} = 10^{10}$ (bottom), demonstrating how network topological properties measurably change as a function of physical parameters--in this case, temperature. The size of the nodes represents the flux-weighted degree of each node. As temperature increases, the weight of the reservoir species increases dramatically, leading to a less uniform distribution of values. These changes in the network as a function of temperature correlate with the change in abundance of CO and $CH_4$ seen below at $K_{zz} = 0$ and $K_{zz} = 10^{10}$, highlighting the transition that occurs at ~1200K.

| Notation | Network Measure |
|---|---|
| $\langle k \rangle$ | Mean degree |
| $\langle k_{nn} \rangle$ | Mean of average neighbor degree |
| $\langle \ell \rangle$ | Average shortest path length |
| $\langle C \rangle$ | Average clustering coefficient |
| $\langle g(v) \rangle$ | Average node betweenness centrality |
| $\langle g(e) \rangle$ | Average edge betweenness centrality |

*Table 1:* Network properties and their notation used in the analysis of the atmospheric chemical reaction networks in this study.

## Thermodynamic Analysis

A number of different thermodynamically-motivated measures of atmospheric disequilibria have been proposed (Line et al. 2014, 2015, 2013, 2012). Among these, the measure developed by Krissansen-Totton et al. (Krissansen-Totton et al. 2016, 2018) has been widely discussed due to its potential applicability to life detection on terrestrial worlds. Here we are interested in including thermodynamic statistics as another component of information in atmospheric models that might be exploited to provide more accurate inferences of relevant planetary properties, such as the degree of disequilibria (e.g. inferred $K_{zz}$). Initially, we measured the thermochemical disequilibria by calculating the total available Gibbs free energy, or $\phi$, for each atmosphere model following Krissansen-Totton et al. (Krissansen-Totten et al 2016). To calculate $\phi$ requires summing the absolute Gibbs free energy of all species at a given temperature, T and pressure, $P_r$:

$$G_{(T,P)} = \sum_i n_i \left(G^\circ_{i(T,P_r)} + RT \ln(a_i)\right) = \sum_i n_i \left(G^\circ_{i(T,P_r)} + RT \ln(Pn_i \gamma_{fi}/n_T)\right) \quad (1)$$

where $G^\circ_{i(T,P_r)}$ is the standard energy of formation for the *i*th species at temperature *T* and pressure $P_r$, *P* is standard pressure (1 bar), and $n_i$ and $\gamma_{fi}$ are the number of moles and the activity coefficient of species *i*, respectively.

In practice, Gibbs free energy is almost always calculated in terms of reference to the standard Gibbs free energy of formation at standard temperature (273.15K) and pressure (1 bar), denoted as $\Delta G_{(T,P)}$. $\Delta G_{(T,P)}$, as Krissansen-Totten et al. 2016 demonstrated, yields the same output as using $G_{(T,P)}$ for the purposes of the minimization required for calculating $\phi$, but is simpler to calculate.

To determine $\phi$, the total Gibbs free energy in Eq. (1) was calculated based on the mole fraction of each species present at equilibrium ($K_{zz} = 0$), and then compared to the mole fraction $n_i$ calculated from simulated atmospheric data for each model in our grid, producing the calculated value of $\phi$ for that model atmosphere, which represents the available Gibbs free energy in the atmosphere:

$$\phi = \Delta G_{(T,P)}(n_i) - \Delta G_{(T,P)}(\overline{n}_i) \quad (2)$$

By construction, $\phi$ as defined by Krissansen-Totton et al. requires an equilibrium value $\phi = 0$. However, equilibrium values cannot readily be established for real-world observations, since many, if not all, planetary atmospheres do not exist at thermochemical equilibrium. Even in our case with a model at equilibrium ($K_{zz} = 0$), observational uncertainty introduced by our interpolating function (see next section) introduces variability in the calculation of $\phi$ that requires a somewhat arbitrary designation of the $\phi = 0$ case as a delta function exactly at $K_{zz} = 0$ and the specified temperature and pressure, from which the difference to interpolated values could be calculated. As the atmospheric model at $K_{zz} = 0$ may or may not reflect the real-world conditions of the atmospheric chemistry in question (or its equilibrium), this assumption introduces additional uncertainty.

This was unlike how we treated the network parameters or chemical species abundances, meaning that we could not make direct comparisons between the utility of $\phi$ and other atmospheric variables. We therefore elected to use $\Delta G_{(T,P)}$, which describes the summed Gibbs free energy of all species in the entire atmosphere, as the primary thermodynamic statistic for each atmosphere, which did not require us to define an equilibrium value for comparison.

## Generating Statistical Distributions via Interpolation of Chemical Species Abundances, Interaction Statistics, and Thermodynamics

To model observational uncertainty the calculated outputs of the VULCAN simulations (species abundances, network statistics, and thermodynamic statistics) were used as the basis of multi-dimensional interpolation functions created using SciPy's griddata package. These functions attempt to 'fit' the estimated position of a data point, in this case, estimated atmospheric model outputs such as species abundances, network measures, and total Gibbs free energy, based on the values of the inputs. Here, the inputs are the initial conditions of temperature and metallicity input into the VULCAN code. That is, when given a set of planetary parameters, for example, $T_{mean}$ = 1200K and $0.5_{solar}$ metallicity, the interpolation function will estimate the most likely values of the physical parameters (such as mixing ratios of specific species) would be if we were to simulate an atmosphere with those conditions. The same process is used for estimating network properties: the output represents what the likely network measurements would be if we had simulated an atmosphere with those conditions, built a CRN based on the resulting abundances of atmospheric gases, and then measured the topological properties of that network. The result is a series of distributions of atmospheric topological measures and physical parameters.

We fed the interpolation functions several 10,000 point normal distributions of initial temperature, metallicity, and pressure values. The centers of these distributions started at 400K, then increased at 100K increments until 2000K, leading to 17 distributions in total.

The likelihood distributions for atmospheric variables can then be statistically analyzed; generally, they were compared on the basis of $K_{zz}$ and temperature (see Figures 5 and 6). To accommodate the limits of certainty that would be present in actual observational data, uncertainties ranging from temperatures of +/- 50K to +/- 1000K were incorporated to generate the distributions. The griddata package cannot interpolate values outside the original range given (i.e., here between 400 and 3000K for mean temperature), therefore many of the conditions drawn from the +/- 1000K Gaussian distribution that was fed into the interpolation function had to be discarded. Consequently, the resulting parameter distributions interpolated from the high uncertainty conditions exhibit a bimodal skew. This is purely an artifact. However, given that this only occurs in situations that already have high uncertainty, and only towards the boundary of the data, it is not expected this will significantly impact the conclusions of our analysis.

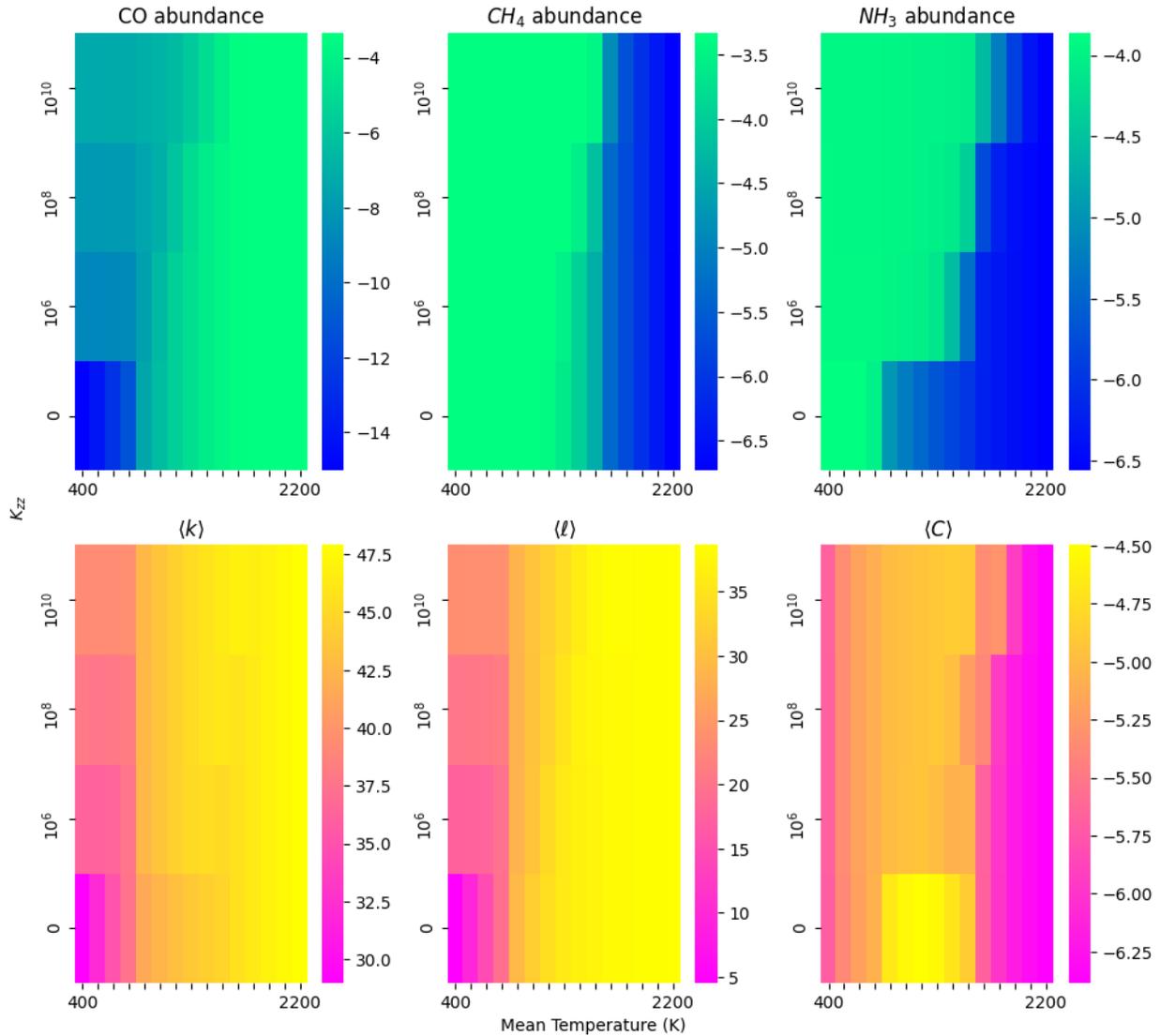

*Figure 4:* *Output from the interpolations functions: heat maps of network measures and species abundance as a function of mean temperature and $K_{zz}$. Metallicity is 1 solar. The x-axis is the mean temperature in Kelvin, the y-axis is $K_{zz}$, and the color scale represents the value of the plotted network measures or abundance. These heatmaps demonstrate that the interpolation function can yield useful insights into relationships between atmospheric parameters. For example, the abundance of CO appears to be positively correlated to high values of mean degree and average shortest path length, whereas $CH_4$ appears to be negatively correlated. $NH_3$'s relationship appears much less clear but does have a weak positive correlation with the average clustering coefficient. These results suggest that it may be possible to infer the network global topology of a planet's atmospheric network from the abundance of certain gases, without necessarily knowing the whole structure of the atmospheric chemical reaction network. For the full set of heat maps detailing our results across multiple metallicity values, see Figures S1-S10 in the Supplement. For the description of the network measures in the figure, see Table 1 and section "Quantifying the Characteristic Topology of Chemical Reaction Networks".*

A key goal of our analyses was to determine if the likely value for a given variable can be statistically distinguished for different $K_{zz}$. This is a necessary precondition for a variable to be a good predictor of $K_{zz}$: it must first take on statistically distinguishable values correlated to $K_{zz}$. This implies the likelihood distributions must themselves be distinguishable for $K_{zz}$ across varying temperatures and uncertainty. We therefore measured the distance between distributions using the Kolmogorov-Smirnoff test, or KS test (Massey 1951)). The two-sample KS test is a nonparametric test useful for distinguishing between two different samples, both with respect to the location and shape of the samples' cumulative distribution functions. The difference is given as

$$D_{n,m} = \sup_x | F_{1,n}(x) - F_{2,m}(x) \qquad (3)$$

Where $F_{1,n}(x)$ and $F_{2,m}(x)$ are the empirical distribution functions of the samples, and $\sup_x$ is the supremum function. Large KS values imply distinguishable distributions, which provide targets for follow-up as possible candidates for accurate inference of $K_{zz}$. Our results show that likelihood distributions of network measures, corresponding to different $K_{zz}$, are distinguishable with respect to physical parameters such as temperature, see Table 2. The distributions of average clustering coefficient were the most distinguishable from each other as a function of $K_{zz}$, followed by mean degree and average shortest path length. This holds true even as the uncertainty of the values increases up to +/- 250K (Figures 5 and 6).

This distinguishability by the use of the KS tests (See Table 2) yields the average clustering coefficient with the highest value across all $K_{zz}$ values, followed by mean degree and average shortest path length. The higher KS scores of these measurements are likely a result of their sensitivity to the reaction rates incorporated into the edge weighting of the network. Other network measurements, such as node betweenness centrality and average neighbor degree, do not track $K_{zz}$ as well.

Compared to measurable molecule abundances, $NH_3$ appeared to be the most distinguishable, having a slightly higher KS value than the network measurements, followed by CO. The other molecule abundances analyzed are only weakly distinguishable from each other as a function of $K_{zz}$, likely due to the fact that the abundance of these species is not significantly altered at different $K_{zz}$ values.

The thermodynamic measurements, $\phi$ and $\Delta G_{T,P}$, are less promising when evaluated using the KS metric. The former yield KS = 1.0 (maximally distinguishable) across all $K_{zz}$ values. However, this is an artifact due to the use of a delta function as a reference equilibrium value (one reason we did not in the end use $\phi$ as the primary thermodynamic measure in this work, though see Figures

S11 and S12, Supplement). The latter actually *decreases* in distinguishability from the equilibrium case as $K_{zz}$ increases.

| $K_{zz}$ | $10^6$ | $10^8$ | $10^{10}$ |
|---|---|---|---|
| **Parameter distributions** | | | |
| *Observable abundances* | | | |
| $NH_3$ | 0.329 | 0.400 | 0.457 |
| CO | 0.266 | 0.294 | 0.320 |
| $CH_4$ | 0.048 | 0.101 | 0.129 |
| $H_2O$ | 0.039 | 0.090 | 0.121 |
| *Network topological measures* | | | |
| Average clustering coefficient | 0.404 | 0.404 | 0.399 |
| Mean average neighbor degree | 0.171 | 0.154 | 0.176 |
| Mean node betweenness centrality | 0.216 | 0.230 | 0.227 |
| Mean edge betweenness centrality | 0.122 | 0.051 | 0.112 |
| Mean degree | 0.165 | 0.195 | 0.235 |
| Average shortest path length | 0.175 | 0.208 | 0.253 |
| *Disequilibrium measure* | | | |
| $\phi$ | 1.0 | 1.0 | 1.0 |
| $\Delta G_{T,P}$ | 0.941 | 0.609 | 0.132 |

*Table 2: Kolmogorov-Smirnoff (KS) metric values for the distributions of observable molecule abundance, network topological measures, and $\phi$. Distributions were calculated from a normal distribution of initial conditions, centered at 900K, $50_{sol}$ metallicity, and 100mb. Values were calculated with respect to the distribution at equilibrium ($K_{zz}$ = 0), except for $\phi$, which was calculated with respect to a 10,000 point delta function set at zero. Uncertainty for the distributions was +/- 50K. A KS value of zero indicates that the two distributions sampled are indistinguishable from each other; the closer the value is to one, the more distinguishable the distributions are. For comparative values across $K_{zz}$ values, see Table S1, Supplement. For the description of the network measures in the figure, see Table 1 and section "Quantifying the Characteristic Topology of Chemical Reaction Networks".*

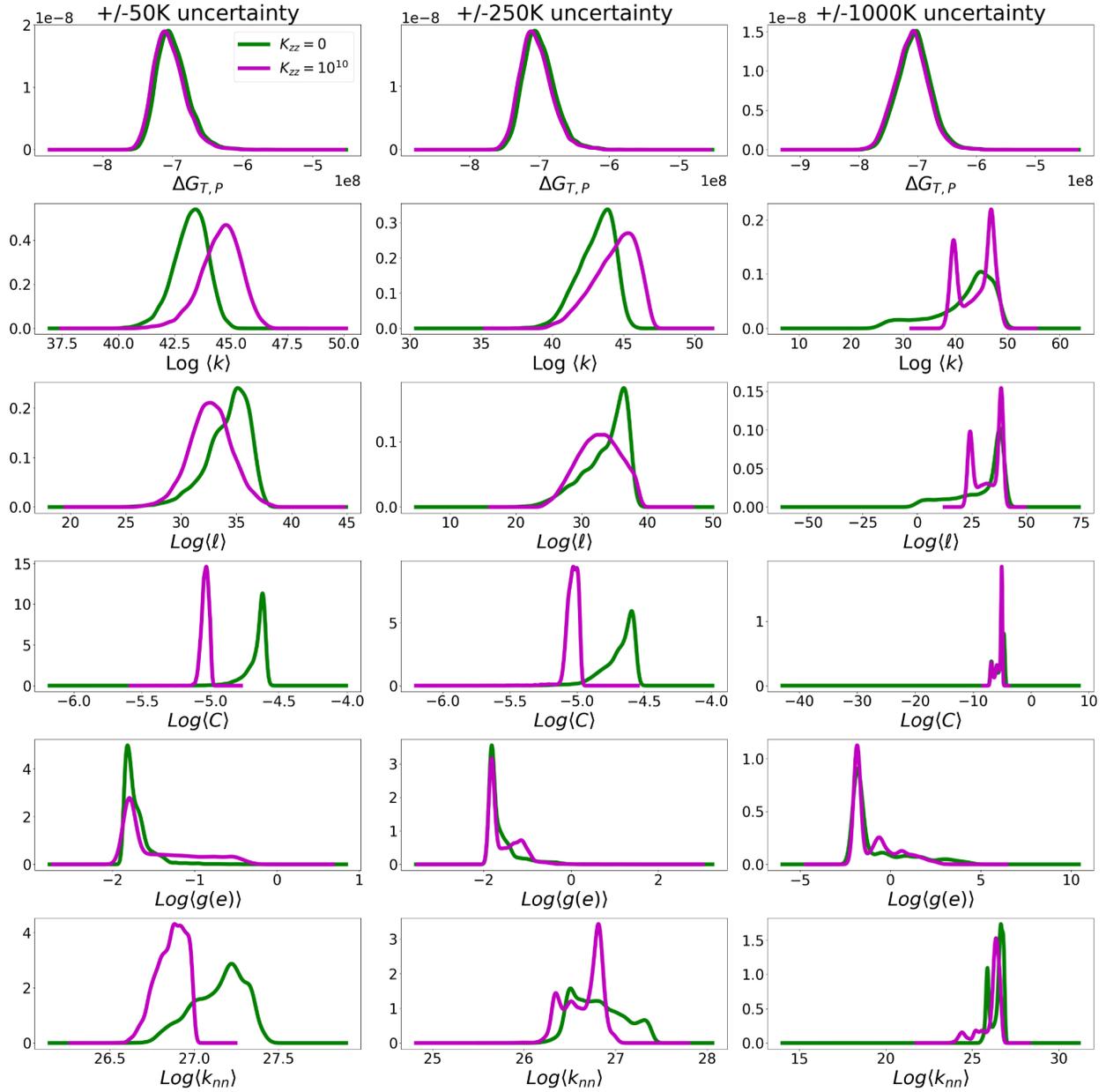

*Figure 5: T= 1200K. Distributions of network measures and thermodynamic parameters, interpolated from a 10,000 point normal distribution of initial conditions centered on a mean temperature of 1200K, at $K_{zz}$ values of 0 and $10^{10}$, and uncertainties in temperature of 50K, 250K, and 1000K. The distributions with different $K_{zz}$ values can be distinguished from each other from the network measures (especially average clustering coefficient and mean degree), even with uncertainties of +/- 250K. Total available Gibbs free energy, however, does not appear to be useful for distinguishing between atmospheres at equilibrium and disequilibrium. The bimodal distributions seen in the +/- 1000K case are an artifact of adding noise to the data. For distributions for all $K_{zz}$ values measured at 1200K, see Figure S13 in the Supplement. For the*

*description of the network measures in the figure, see Table 1 and section "Quantifying the Characteristic Topology of Chemical Reaction Networks".*

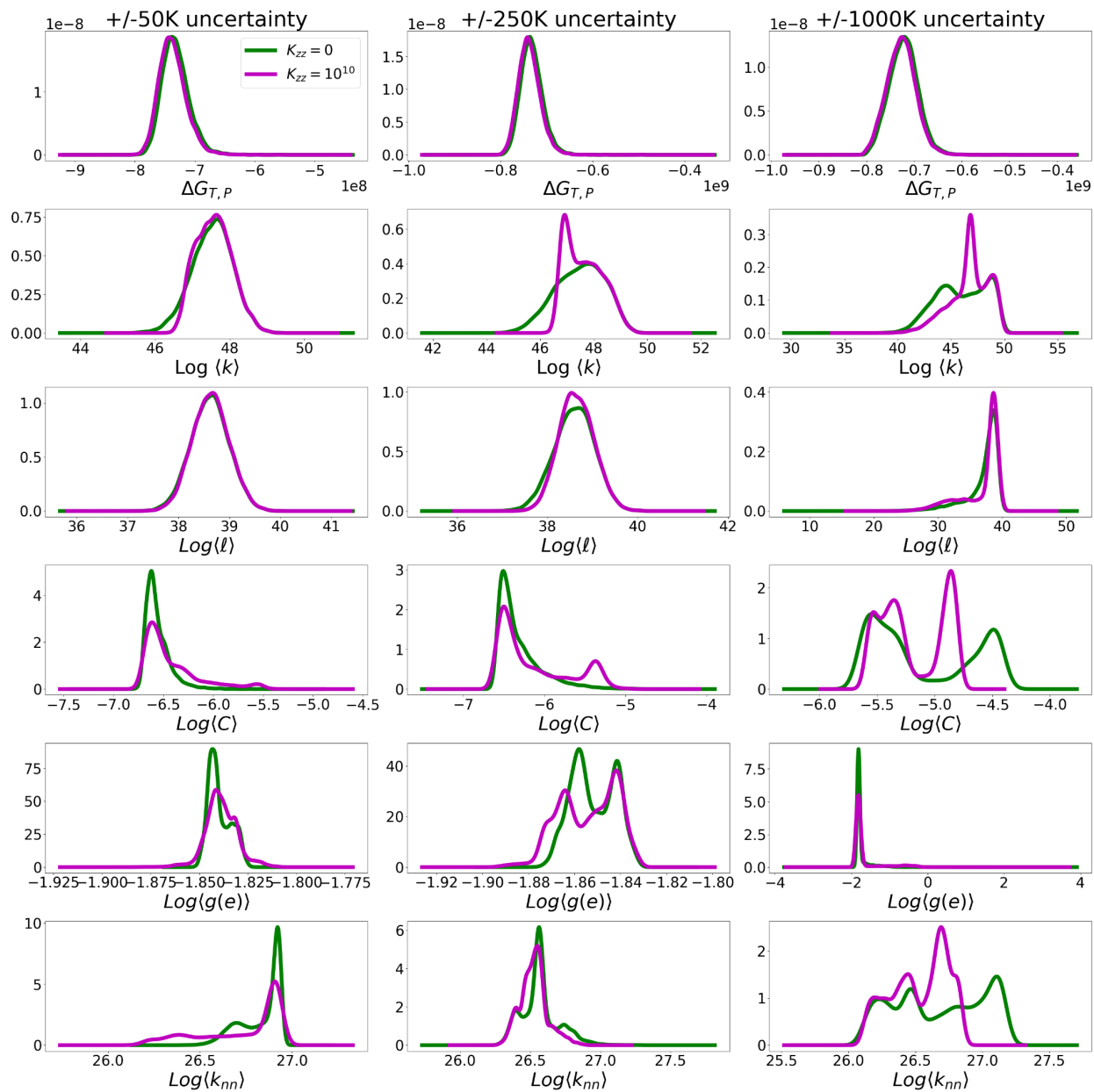

***Figure 6: T= 2000K.*** *Distributions of network measures and molecular abundances, interpolated from a 10,000 point normal distribution of initial conditions centered on a mean temperature of 2000K, at $K_{zz}$ values of 0 and $10^{10}$, and uncertainties in temperature of 50K, 250K, and 1000K. Distributions with different $K_{zz}$ values are less distinguishable from each other, likely due to the faster thermal kinetics forcing the system back to equilibrium more rapidly than can be dredged up via vertical mixing. The bimodal distributions seen in the +/- 1000K case are an artifact of adding noise to the data. For distributions for*

*all $K_{zz}$ values measured at 1200K, see Figure S13 in the Supplement. For the description of the network measures in the figure, see Table 1 and section "Quantifying the Characteristic Topology of Chemical Reaction Networks".*

## Evaluating Model Error

We analyzed the effect of perturbation on the network analysis through several interventions to the network. First, we perturbed the data by removing the $CH_4$-CO quenching pathway (selected due to the critical role it plays in the atmospheric chemistry of hot Jupiters). We also removed all species containing oxygen, carbon, or nitrogen from the network (see Figures S15 and S16, Supplement), either singularly or in combination with another element being removed, both to see how the network topology was changed, and to try to understand the more subtle underlying network features that might have been obscured by the extremely high-weight edges. We also did perturbation testing by removing key species--$O_2$, $CH_4$, $H_2O$, and $NH_3$-- and their associated reactions. Like the uncertainty inserted during interpolation, these tests were conducted to determine the analysis pipeline's tolerance for error (either stemming from observations, or the modeling process, both of which could limit our abilities to make accurate inferences).

We found that perturbing the networks changed the distribution of network measurements, even when only the $CH_4$-CO pathway was removed (see Figure 7). This change in distribution was likely due to the fact that the networks are edge-weighted by reaction rates, and that this pathway has an extremely high edge-weight, especially at high temperatures. Unsurprisingly, the changes were even more pronounced when large swathes of species were removed, such as all carbon-containing compounds (see Figure S15, Supplement).

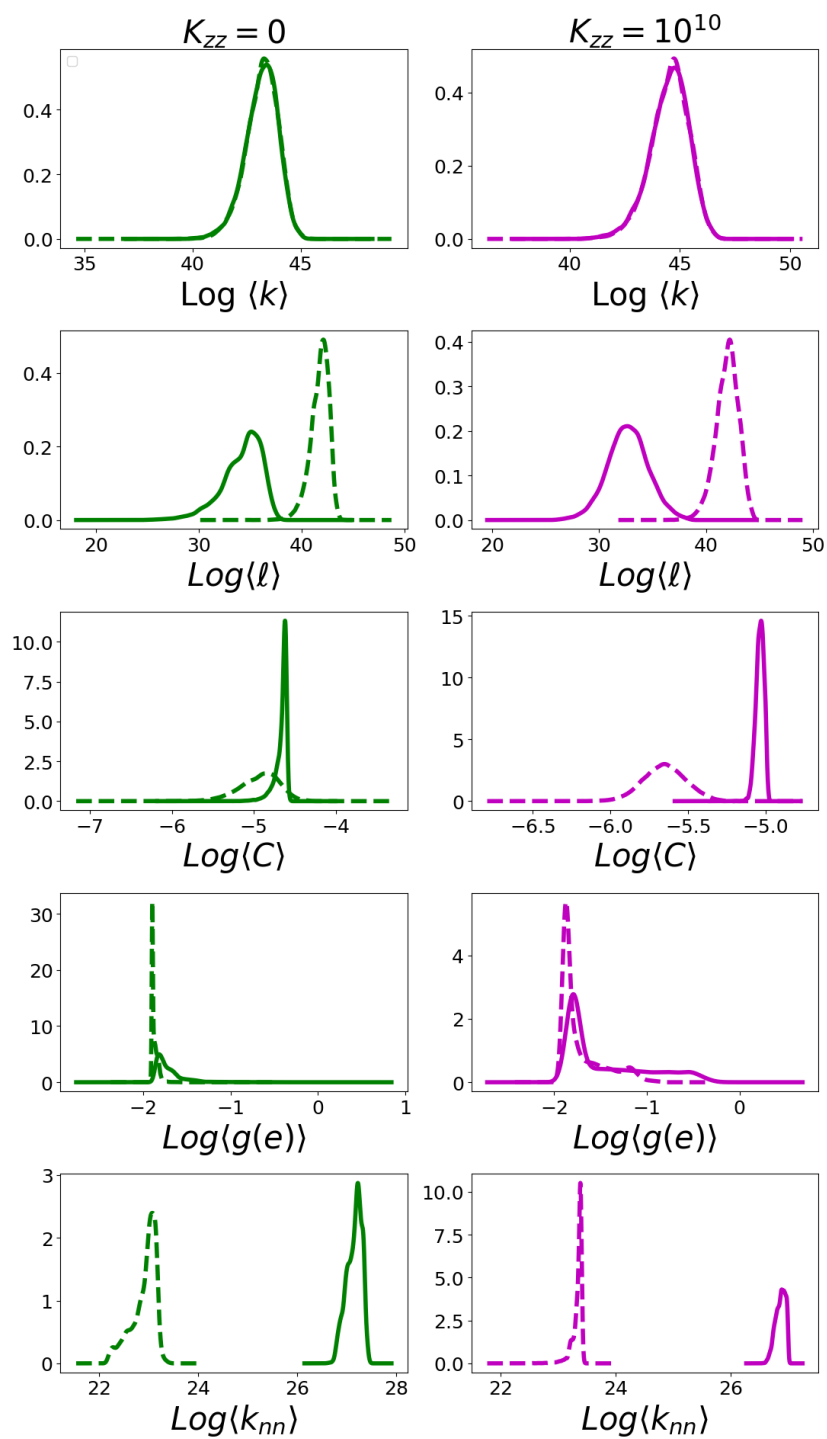

*Figure 7: T=1200K w/ Perturbations w/ pathway removed, uncertainty +/- 50K. Distributions of network measures when the $CH_4$-CO quenching pathway is removed from the network (dashed line) vs complete networks (solid line), at 1200K and 1 solar metallicity. Average shortest path length and betweenness centrality allow distinction between the two. For the description of the network measures in the figure, see Table 1 and section "Quantifying the Characteristic Topology of Chemical Reaction Networks".*

Predictive Efficacy of Multivariate Data for the Disequilibria of Atmospheric Networks.

We next sought to evaluate the predictive power of the multivariate measures to identify the disequilibria of atmospheric chemical networks beyond their distinguishability. To do so, first, we grouped multivariate measures into three groups: (1) thermodynamic measure - Gibbs free energy, (2) network topological measures - mean degree, average shortest path length, average clustering coefficient, mean value of average neighbor degree, average node betweenness centrality, average edge betweenness centrality, (3) abundance of four chemical species - $H_2O$, $CO$, $CH_4$, and $NH_3$. We then classified the dataset with a given set of input variables into four different groups associated with $K_{zz}$ = 0, $10^6$, $10^8$, and $10^{10}$ using XGBoost (Chen & Guestrin 2016; Sharma 2018), a supervised machine learning algorithm, i.e. predicting $K_{zz}$ of the networks as a target variable from the three different kinds of measures (1-3 above), respectively. After that, we evaluate the prediction accuracy from each group of variables. This approach allows us to not assume a functional form for the model but instead rely on data alone for insights. For the actual implementation of XGBoost, we split the dataset into training and testing sets with 80-20 ratio randomly using a Python package *sklearn.model_selection.train_test_split* and train *XGBClassifer* from a Python package, *xgboost* (Chen & He), (Installation Guide — xgboost 1.4.0-SNAPSHOT documentation 2020) on the training sets with min_child_wight=10, max_depth=5, and n_estimators=1000.

The results, see Figure 8, show that the set of global topological measures of the atmospheric networks is the best predictor for $K_{zz}$ throughout datasets with different mean temperatures and uncertainty in temperature. The Gibbs free energy in the figure, G, alone has the lowest prediction accuracy around 70% throughout the mean temperature range from 400K to 2000K regardless of uncertainty in temperature. On the other hand, the abundance of the four chemical species and the topological measures in Figure 8 show high accuracy in the prediction that is higher than or close to 90% from 400K to 1600K, before the accuracy drops after 1800K when the uncertainty in temperature is 50K and 250K, respectively. When the uncertainty of temperature in the dataset is 1000K, even though the accuracy of predicting $K_{zz}$ from the set of network topological measures or the set of abundance decreases as the mean temperature of the model increases, both groups of variables are better predictors of the disequilibria states than Gibbs free energy. Finally, Figure 8 shows that network topological measures outperform abundance measures in most parts of the range of mean temperature regardless of uncertainty in temperature for predicting $K_{zz}$.

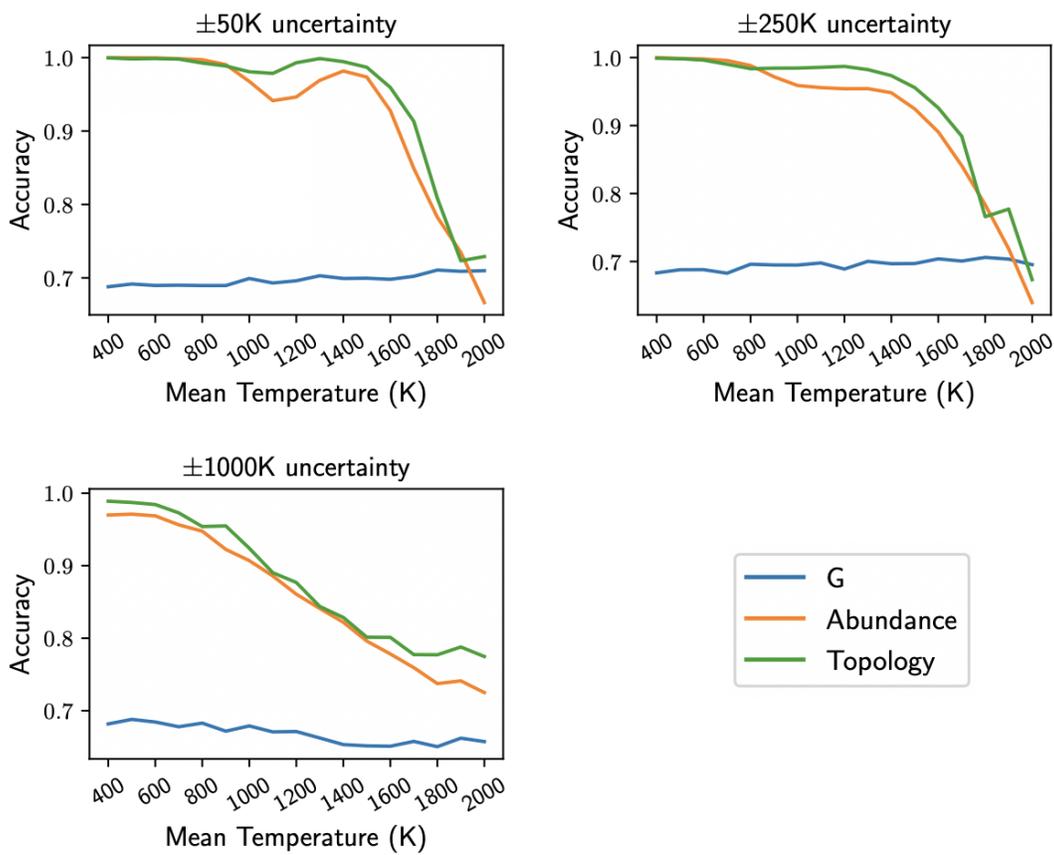

*Figure 8:* *Predicting disequilibrium state of the atmosphere, $K_{zz}$ from the three different sets of variables: G (blue), Abundance (orange), and Topology (green). G indicates Gibbs free energy, abundance the abundances of chemical species $NH_3$, $CH_4$, CO, and $H_2O$, and the average topological properties are those listed in Table 1. For the description of the network measures in the figure, see section "Quantifying the Characteristic Topology of Chemical Reaction Networks". G alone had the lowest accuracy; topology alone had the highest accuracy at lower mean temperatures but decreased accuracy as mean temperatures rose.*

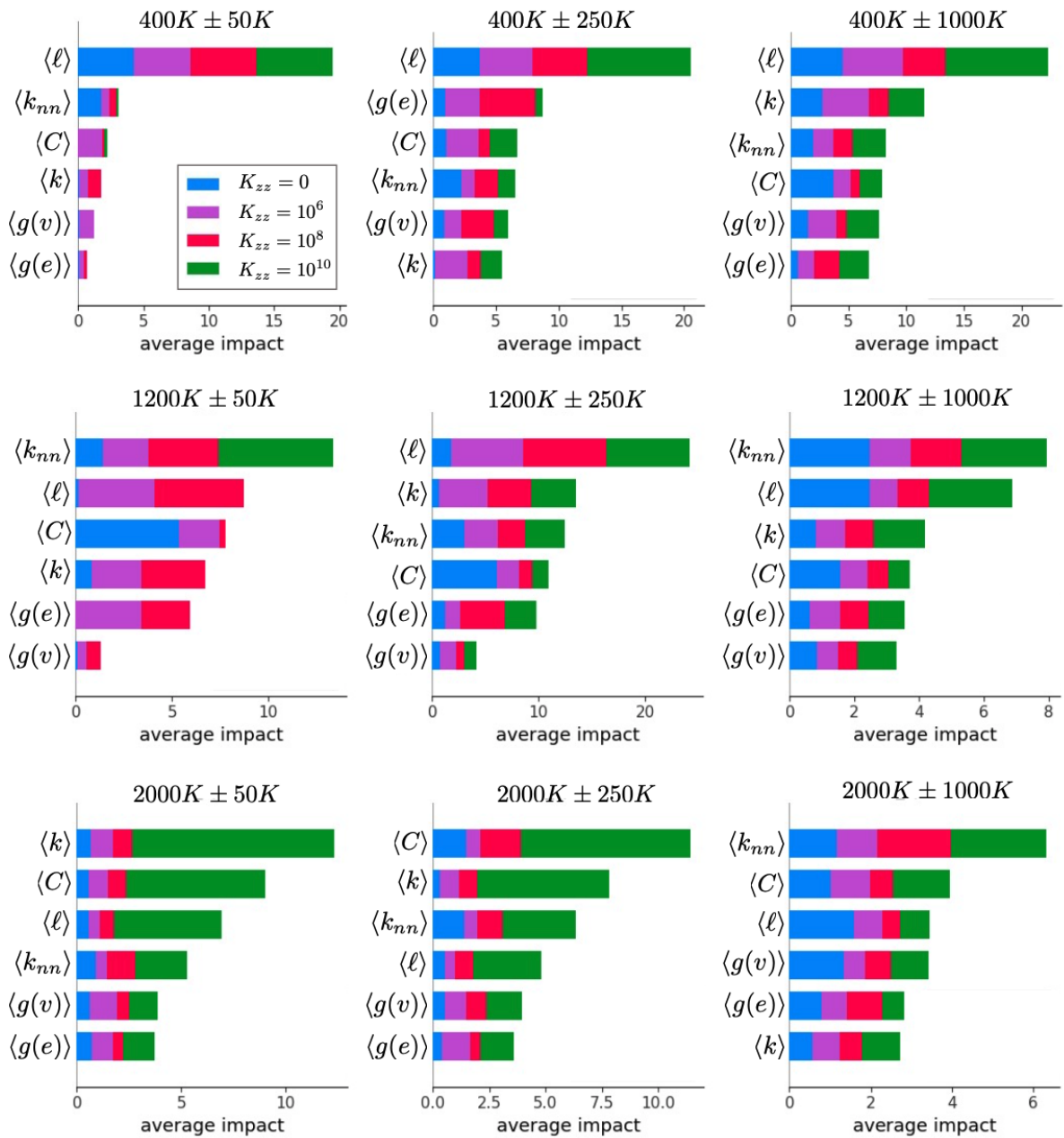

***Figure 9:*** *The mean SHAP values of network topological measures in Table 1, over the range of $K_{zz}$ values. The size of bars represents the contribution of topological measures included to the prediction for different $K_{zz}$. For the mean temperature of 400K, the average shortest path length causes 2-10 times more change in the prediction. The difference amongst the influence of topological measures decreases as the mean temperature or the uncertainty of temperature increases. This observation suggests that multivariate analysis utilizing various topological measures has the advantage of characterizing the disequilibria of the reaction networks when compared to attempting to find a single characteristic measure valid over a wide range of temperature with different levels of uncertainty in the data.*

## Evaluating the Importance of Network Topology for Predicting the Disequilibria of Atmospheric Networks.

After demonstrating that network topological measures are the best group predictor for $K_{zz}$ among other variables, our next natural question is how much the individual network topological measure contributes to the prediction. To this end, we use the SHAP (SHapley Additive exPlanations) values (Lundberg S. M. 2017), which is based on a model-agnostic interpretation method for machine learning prediction, Shapley values from cooperative game theory (Shapley 1951). To determine the SHAP value a game is played with all possible combinations of variables as the players, and with the payout being the difference of the predicted value from the marginalized value. Let $f$ be the predictive model or classifier built from XGBoost in the previous section. The SHAP value for a variable, $x$ in the classification is measured from the difference between a value function that includes the variable, $x$, and the value function which is marginalized over $x$. Using the additive property of Shapley values, the average influence of the variable, $x$ is determined by the ensemble average of the absolute SHAP values. The higher the average absolute SHAP value, the more important is the variable to predict $K_{zz}$ (see Figure 9).

## Robustness of Network Topology as Predictor of the Disequilibria of Atmospheric Networks under Perturbation

As mentioned above, perturbed datasets were created by removing species or pathways during the process of generating the weighted network, and then using these incomplete networks as the basis for the interpolation function. To test the robustness of network topological measures, we also produced networks where one of four key species--$NH_3$, $CH_4$, $CO$, and $H_2O$--was removed during the edge weighting of the network. As before, these networks were otherwise analyzed identically to the non-perturbed networks.

We implemented the same XGBoost machine learning algorithm for predicting $K_{zz}$ based on the perturbed network datasets with uncertainty 50K. As shown in Figure 10, the set of topological measures is the best predictor for $K_{zz}$ compared to Gibbs free energy or the set of the abundance of chemical species, just as for the non-perturbed datasets. The prediction accuracy from the topological measures show notable changes before and after the perturbation only at high mean temperature, 1600K ~ 2000K: after removal of $CH_4$, $CO$, and $H_2O$, respectively, the prediction accuracy increases compared to the unperturbed datasets while after removal of $NH_3$ caused decrease in the accuracy.

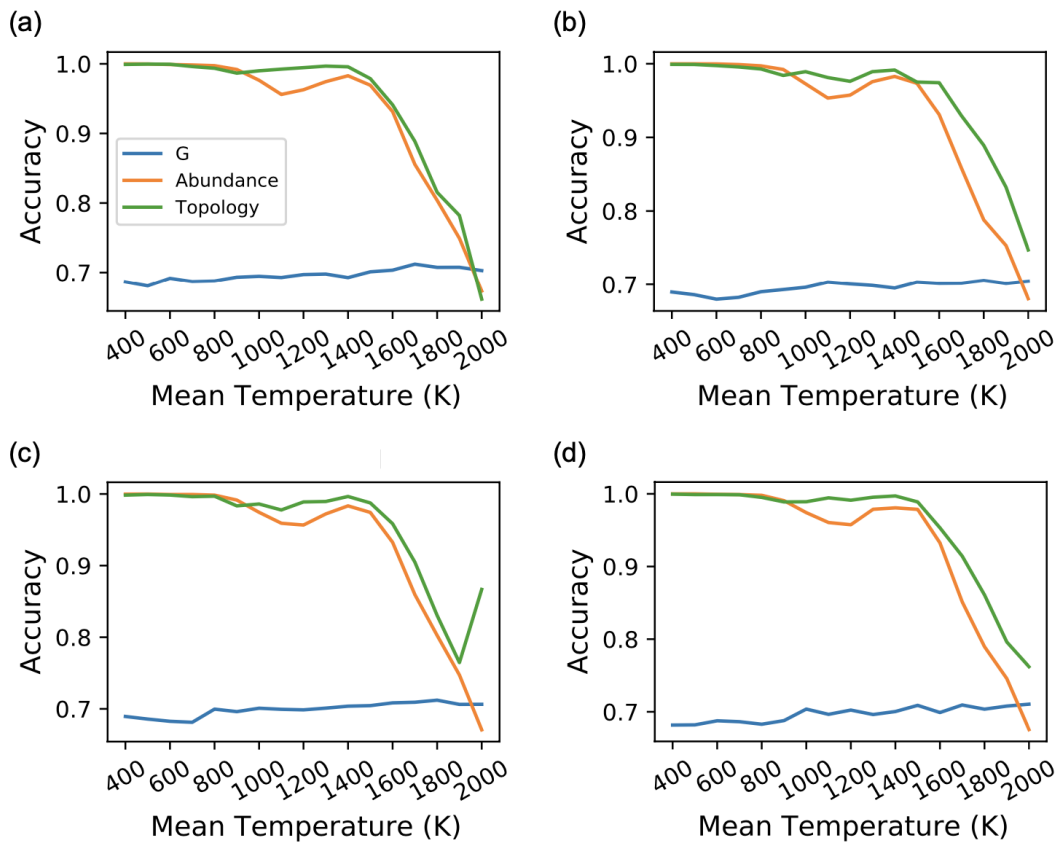

*Figure 10:* *Robustness of topological measures (in Table 1) as a predictor of $K_{zz}$: accuracy of machine learning to predict $K_{zz}$ of the perturbed reaction networks where one of four chemical species; (a) $NH_3$, (b) $CH_4$, (c) CO, and (d) $H_2O$, was knocked out. Predictions based on the thermodynamic measure, G (blue) shows poor performance compared to topological (orange) and abundance (green) measures. The uncertainty of temperature in the given dataset is 50K.*

## Discussion

Current approaches to infer the properties of exoplanets from their atmospheres aim to identify specific molecular species from spectral data as the first point of contact. From these data the most likely models can be fit, allowing indirect inference of properties not directly observable, such as vertical mixing strength. In the current work, we have expanded the types of variables exoplanet researchers should consider when building new methods for remote inference. Thermodynamic statistics and network topology can be derived from the same models used to predict species abundances, but provide different information about the chemistry happening within a planetary atmosphere.

Our results indicate that topological measurements can be used to distinguish between atmospheres with different $K_{zz}$ values, and thus, correspondingly different distances from thermochemical equilibrium. Furthermore, topology displays greater discriminating power than abundances alone in cases where we have complete information about the atmospheric chemistry of the planet. This is because, in constructing the weighted edges of the network, we are able to incorporate information regarding the reaction kinetics that would not be available from direct observations of the atmosphere.

This holds true in terms of machine learning analysis and predictability, as well. As long as the data is not missing central species, topology offers higher accuracy as a basis for machine learning algorithms than abundance across a wide range of temperatures. Furthermore, even in cases where critical species are missing, topology-trained algorithms offer performance that's still as accurate as those trained on abundances alone.

These findings appear resilient even in the face of considerable observational uncertainty, as the distributions of average network clustering coefficient, for example, remain distinguishable even with uncertainty in temperature of +/- 250K. To a lesser extent, we can also distinguish between systems that have been perturbed via the removal of critical pathways.

Individually, it is clear that no one measure will single-handedly provide confidence in the inference of disequilibria given the observational and model uncertainty, and instead, our results demonstrate how a multivariate approach is necessary. Average clustering coefficient, for example, demonstrates a strong ability to determine if the atmosphere is in a state of equilibrium or not, but lacks the sensitivity to indicate how far away the system is from equilibrium. Mean degree and average shortest path length, on the other hand, provide much greater sensitivity to the distance from equilibrium, even if they are less sensitive overall compared to average clustering coefficient. Abundances and thermodynamic measurements can provide grosser distinctions but at higher sensitivities.

A caveat of our approach is that network statistics are calculated from prior knowledge of the abundances and rates. One can therefore reasonably raise the question of what advantage an approach based on network topology really brings if it requires information already captured to a lesser degree in other variables, e.g., individual species abundances (which are statistically combined to yield weighted network topology). There are two key reasons that the network-based approach provides utility beyond just knowing abundances. The first is that we have shown that network topology is more predictive of atmospheric disequilibria than abundances alone. This means that additional information about the degree of disequilibria is captured in system-level patterns in the reactions happening in an atmosphere, which can be quantified by representing the reactions as a complex network. Thus, even if network representations of atmospheric chemistry are not ultimately adopted as a tool for remote inference, they do provide utility in understanding disequilibria properties of atmospheres. The second reason is that our work motivates why network topology should be a target for new inference methods, in particular by targeting direct inference of network statistics from atmospheric data rather than via abundances. If this can be done it allows a more direct means of getting at features (e.g. topology) that most accurately predict disequilibria from spectral data.

This latter point feeds into a major part of our motivation for studying atmospheric network properties alongside more traditional variables for inference such as abundances. This is due to their potential for future development of statistically motivated methods for the remote inference of life. Biosignatures based on statistics of molecular interactions in the atmospheres of exoplanets have been proposed by a number of researchers as a promising candidate for future life detection efforts (Centler & Dittrich 2007; Estrada 2012; Schwieterman et al. 2018; Walker et al. 2020), The reasoning is multifold, but relies on the idea that life is a systems-level property and that when detecting life we need agnostic, quantitative approaches that aim to measure the presence of life at the level of interacting systems. To accomplish this goal, two near-term tasks are required (1) network structure must be validated as a statistically significant distinguisher of a biosphere on a planet, and (2) it must be demonstrated that network properties are remotely inferable. Efforts to address (1) are underway (Kim et al. 2018), and there is ample evidence that biologically driven networks display distinctive topological properties. Here, we have attempted to take the first steps toward (2) that are rigorously grounded in the realities of remote detection. Our goal was to demonstrate that useful information about planetary atmospheres is contained in their reaction network topology, and indeed we find this is the case. Taken in a multivariate analysis, network topology provides additional statistical power for inferring the state of atmospheric disequilibria. In doing so, we have demonstrated that network statistics can be a useful tool for exoplanet characterization, even outside of life detection. However, much remains to be done to develop these methods further and to make them a powerful statistical tool for the exoplanet science community. One future direction of utility is to develop methods for directly inferring network structure from atmospheric spectra, which could allow more robust inferences at the level of molecular interaction statistics rather than of individual chemical species when studying planetary atmospheres. Furthermore, such methods would allow directly inferring features relevant to life if the approach of (1) is successful in statistically discriminating it.

## Acknowledgements

We gratefully acknowledge support from the National Aeronautics and Space Administration via NASA's NExSS grant NNX15AD53G.

# Supplementary Materials

# Inferring Disequilibria in Exoplanet Atmospheres using Multivariate Information

Theresa Fisher[1], Hyunju Kim[1,2,3], Camerian Millsaps[1], Michael Line[1], Sara Walker[1,2,3,4*]

[1]School of Earth and Space Exploration, Arizona State University, Tempe AZ USA
[2]Beyond Center for Fundamental Concepts in Science, Arizona State University, Tempe AZ USA
[3]ASU-SFI Center for Biosocial Complex Systems, Arizona State University, Tempe AZ USA
[4]Santa Fe Institute, Santa Fe, NM USA
*author for correspondence: sara.i.walker@asu.edu

# Methods

## Atmospheric Modeling with VULCAN

The network analysis was a multi-step process, using several different techniques. The first step of the process was to use the VULCAN modeling package (Tsai et al 2017) to simulate the atmospheric chemistry of a hot jupiter over a range of temperatures (400K to 3000K, in ~100K increments) and, later, metallicities (0.1 to 300 times that of the solar system, evaluated on a log scale) and pressures (from 50mb to 150mb, reflecting the likely depth in the atmosphere that will be observable), and finally, vertical mixing coefficients, ranging from zero to $10^{10}$ s$^{-1}$ cm$^{-1}$.

The vertical mixing coefficient, or $K_{zz}$, represents how much mixing is occurring between the lower layers of the atmosphere, which are at higher temperature and pressure, and the upper layers, which are at lower temperature and pressure. At a $K_{zz}$ value of zero, there is no mixing between layers, and each layer is assumed to very quickly reach chemical equilibrium; thus, we consider this to be our equilibrium case. As the atmosphere becomes more mixed, it moves away from equilibrium, since molecules can be produced at the lower levels of the atmosphere in quantities that are not thermodynamically favorable at the higher levels and then "dredged" up to the higher levels at a rate faster than they can be depleted by equilibrium chemistry. Thus, $K_{zz}$ is used as a proxy of how far away the atmospheric chemistry is from equilibrium.

To compare different types of disequilibrium, model simulations were also conducted where photochemical reactions were included in addition to non-zero vertical mixing coefficients.

## Network Analysis

Once the models converged to a solution, the results were saved as a .vul file, a format specific to VULCAN that can be treated as a Python dictionary once unpacked with the pickle function. Next, the chemical reaction network is created using a pre-existing reaction list derived from the one used in the source code of VULCAN. The abundances of each constituent component of the atmosphere, and the rate constants of each reaction, were extracted using a Python pipeline and used to weight the edges between nodes.

The topological properties of these chemical reaction networks were then measured using the Python NetworkX package. Specifically, after ensuring the network graphs were made up of the largest connected components, we used the degree, clustering, betweenness_centrality, average_shortest_path_length, and average_neighbor_degree.

The network measurements and abundances, along with initial conditions (in terms of temperature, pressure, and metallicity), were used as the basis of multi-dimensional interpolation functions created using SciPy's griddata package. These functions can then be used to estimate a value for the network parameters based on an initial temperature and metallicity.

## Thermodynamic Analysis

To compare our findings to existing thermochemical disquilibrium metrics, we also calculated the total available Gibbs free energy, or $\phi$, for each atmosphere modeled (Krissansen-Totten et al 2016). We used a version of the code provided by Krissansen-Totten et al (2016) implemented in Python for consistency.

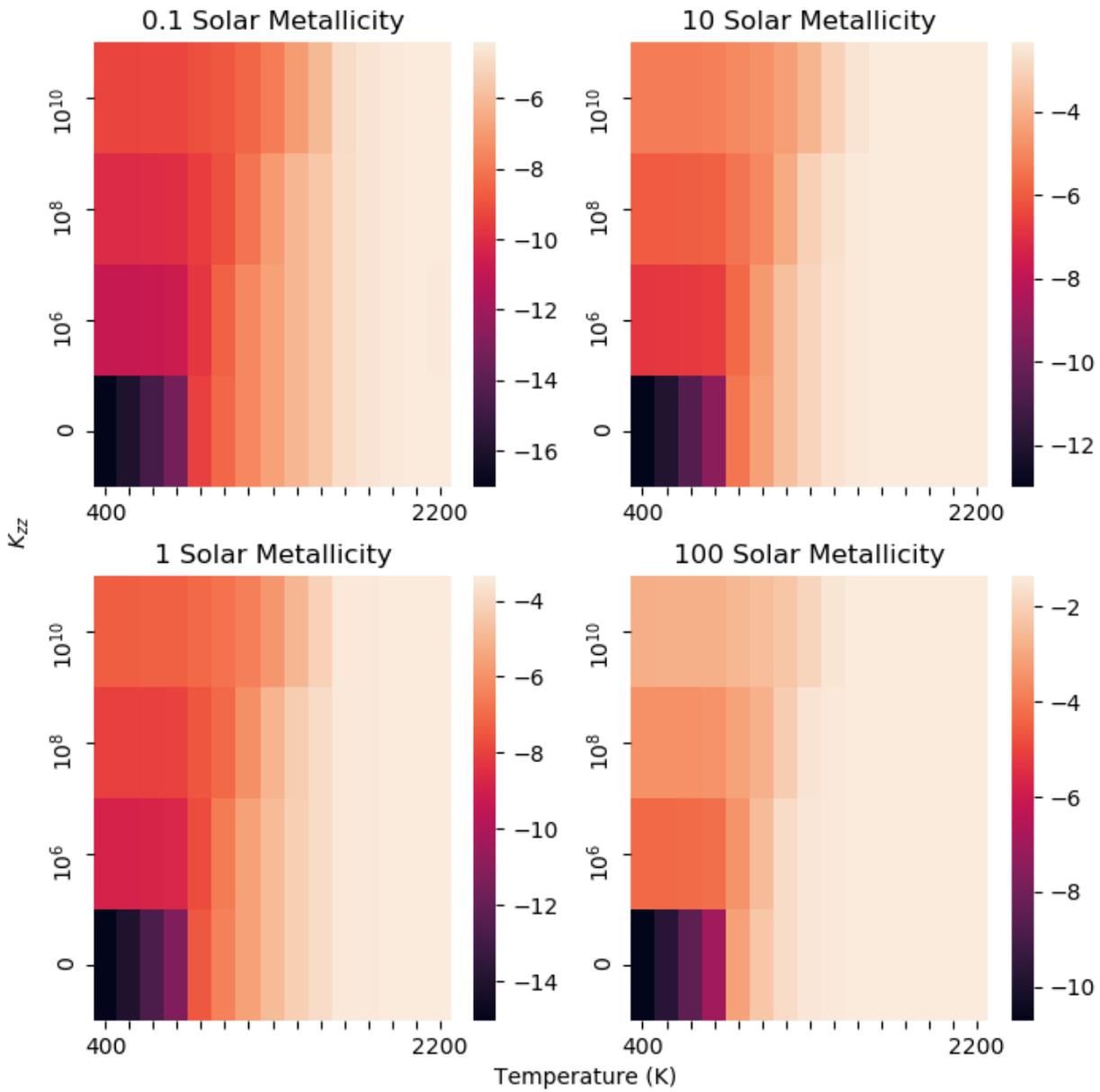

***Figure S1:*** *The abundance of CO as a function of temperature and $K_{zz}$, at 0.1, 1, 10, and 100 times solar metallicity. X-axis is temperature, y-axis is $K_{zz}$, and color corresponds to the abundance of CO.*

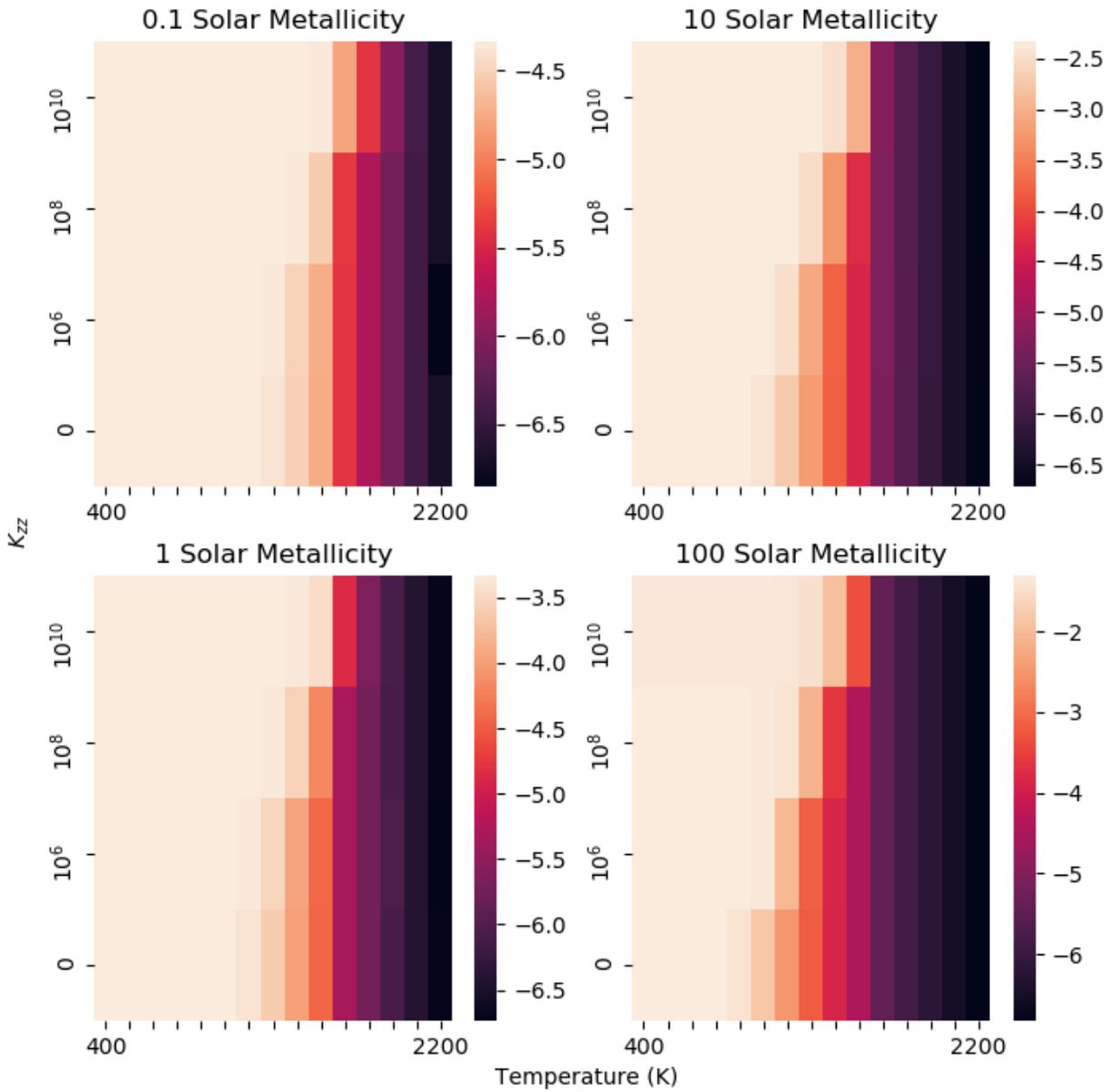

***Figure S2:*** *The abundance of $CH_4$ as a function of temperature and $K_{zz}$, at 0.1, 1, 10, and 100 times solar metallicity. X-axis is temperature, y-axis is $K_{zz}$, and color corresponds to the abundance of $CH_4$.*

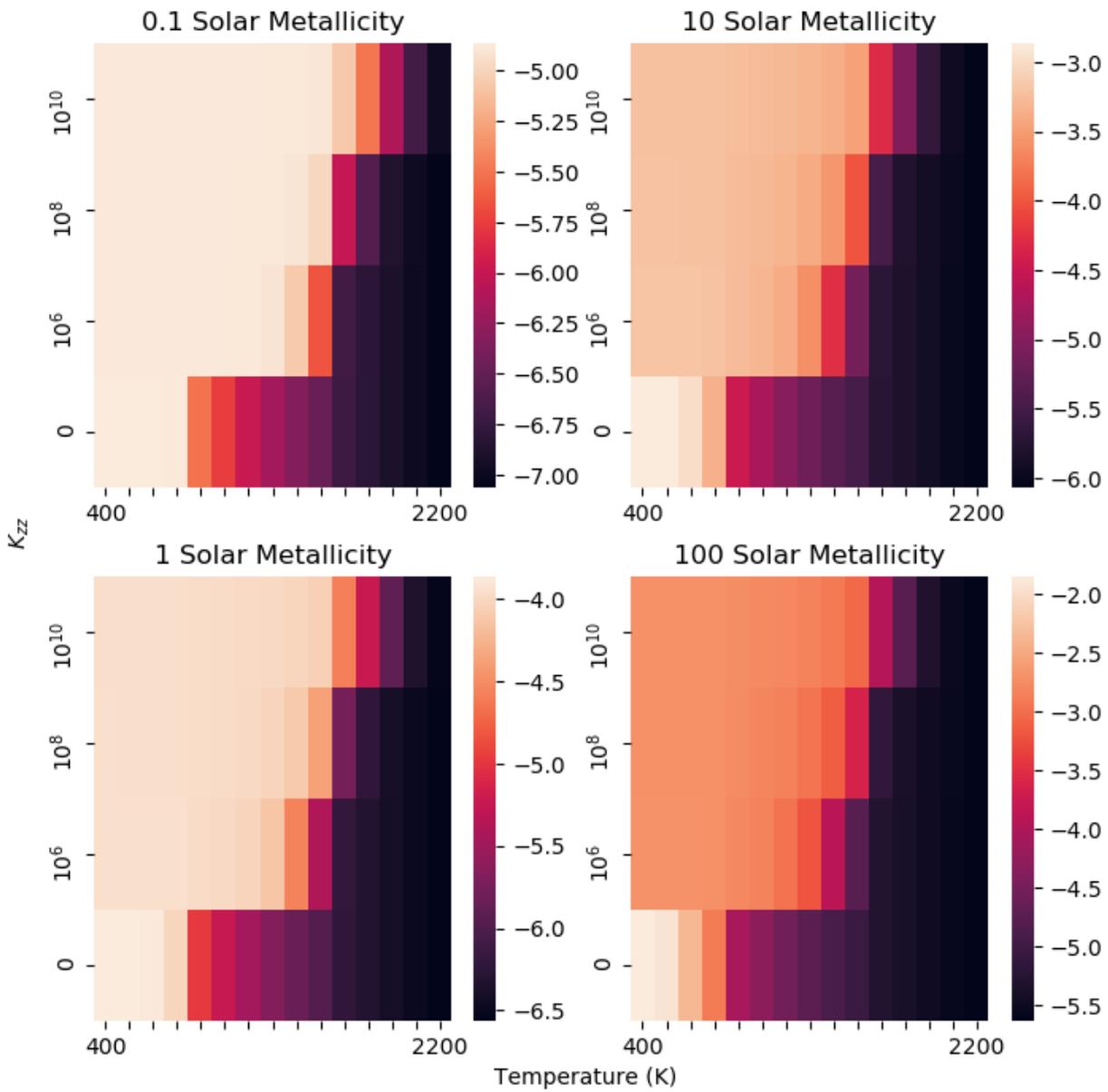

***Figure S3:*** *The abundance of NH$_3$ as a function of temperature and K$_{zz}$, at 0.1, 1, 10, and 100 times solar metallicity. X-axis is temperature, y-axis is K$_{zz}$, and color corresponds to the abundance of NH$_3$.*

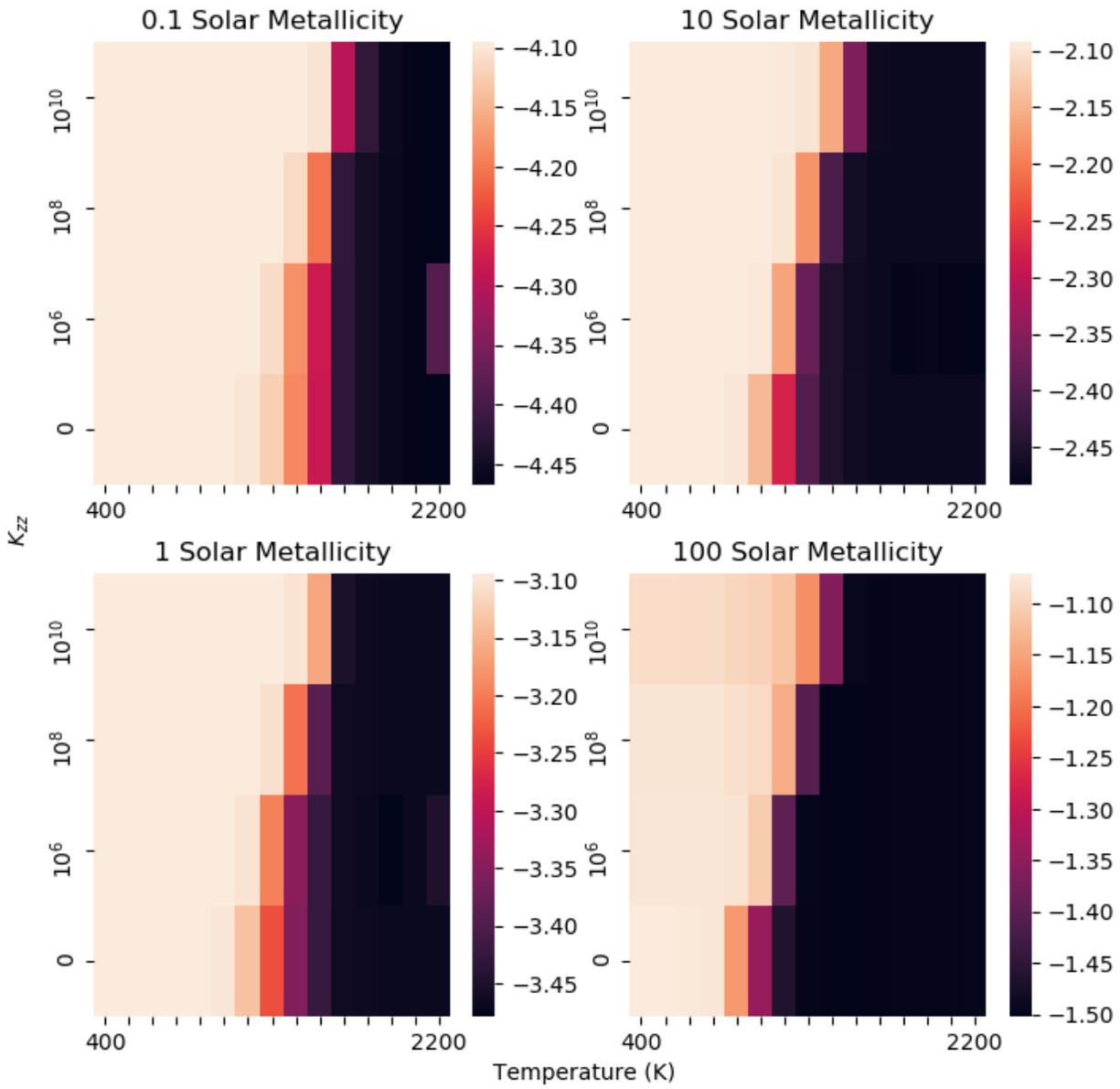

*Figure S4:* *The abundance of $H_2O$ as a function of temperature and $K_{zz}$, at 0.1, 1, 10, and 100 times solar metallicity. X-axis is temperature, y-axis is $K_{zz}$, and color corresponds to the abundance of $H_2O$.*

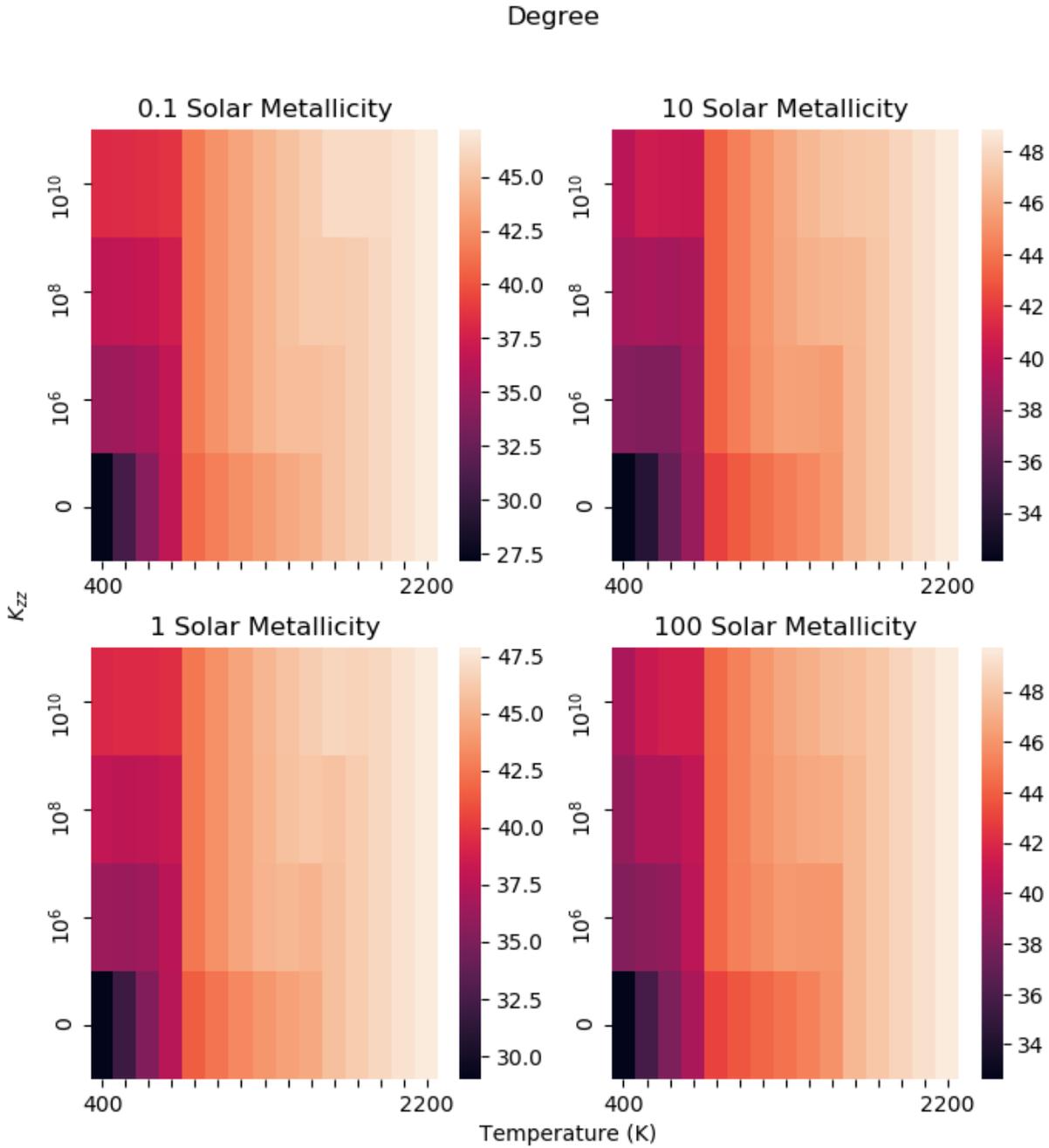

*Figure S5:* Mean degree a function of temperature and $K_{zz}$, at 0.1, 1, 10, and 100 times solar metallicity. X-axis is temperature, y-axis is $K_{zz}$, and color corresponds to the weighted mean degree.

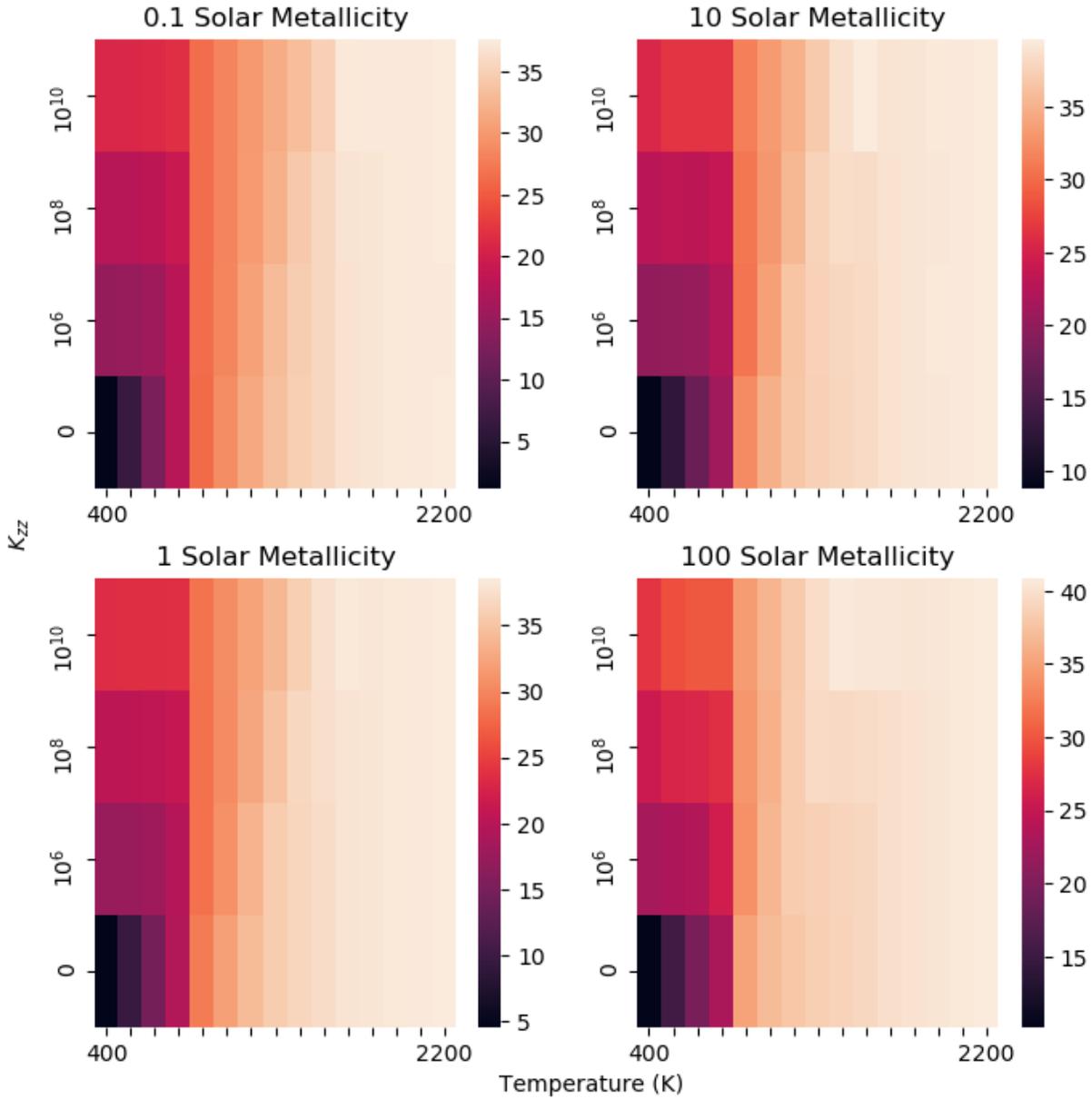

*Figure S6:* Average shortest path length as a function of temperature and $K_{zz}$, at 0.1, 1, 10, and 100 times solar metallicity. X-axis is temperature, y-axis is $K_{zz}$, and color corresponds to the weighted average shortest path length.

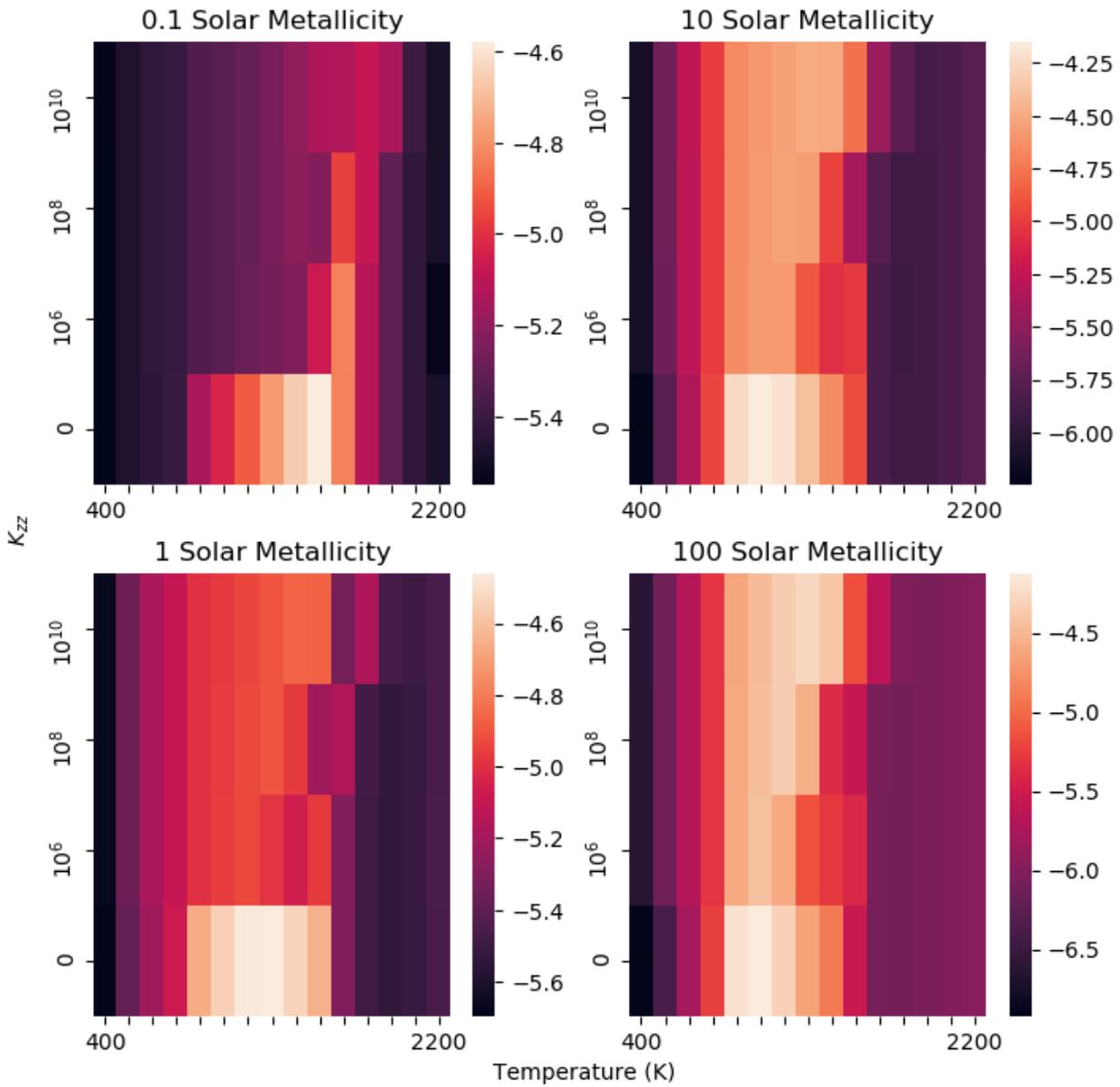

*Figure S7:* *Average clustering coefficient as a function of temperature and $K_{zz}$, at 0.1, 1, 10, and 100 times solar metallicity. X-axis is temperature, y-axis is $K_{zz}$, and color corresponds to the weighted average clustering coefficient.*

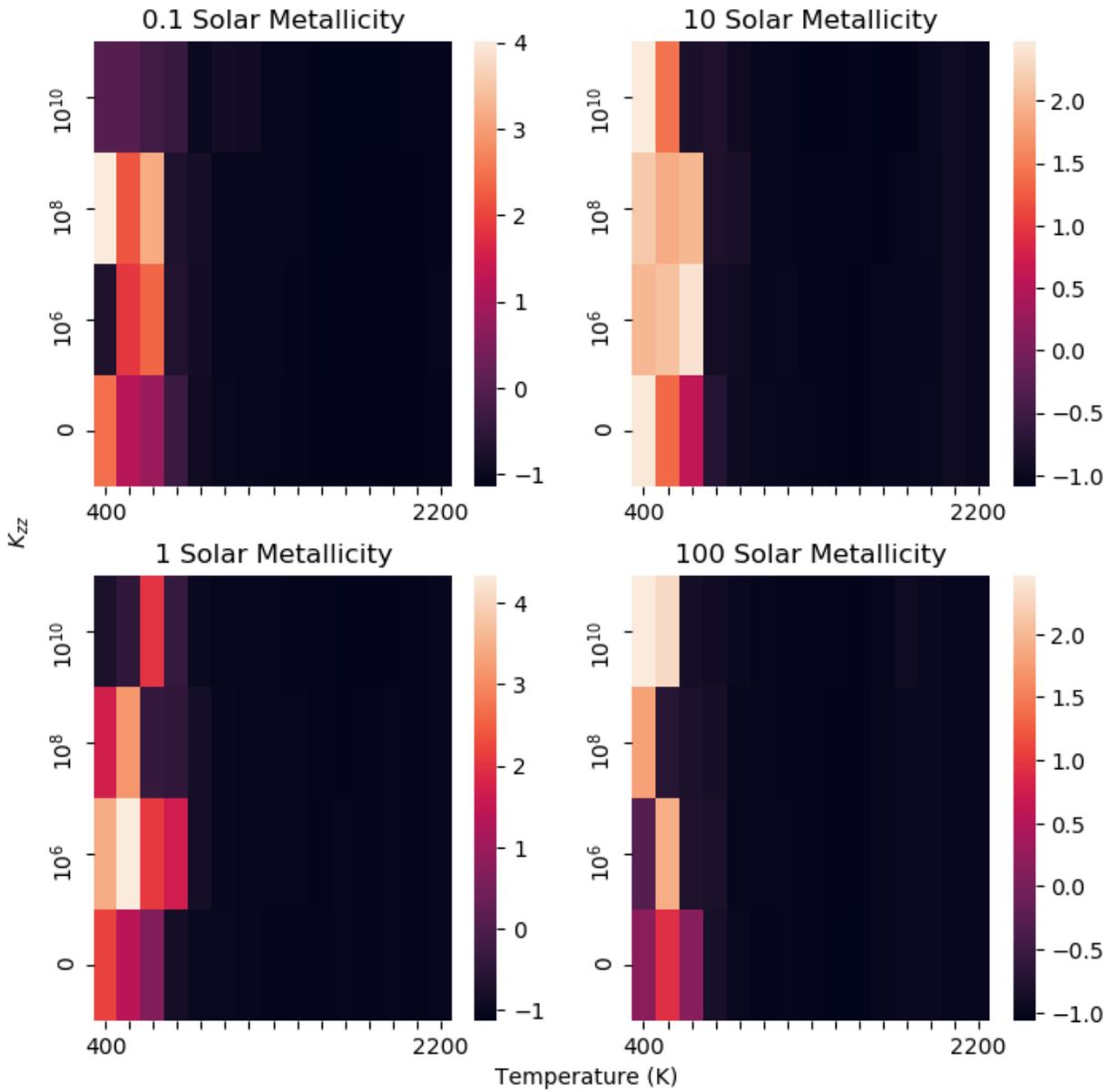

***Figure S8:*** *Node betweenness centrality as a function of temperature and $K_{zz}$, at 0.1, 1, 10, and 100 times solar metallicity. X-axis is temperature, y-axis is $K_{zz}$, and color corresponds to the weighted node betweenness centrality.*

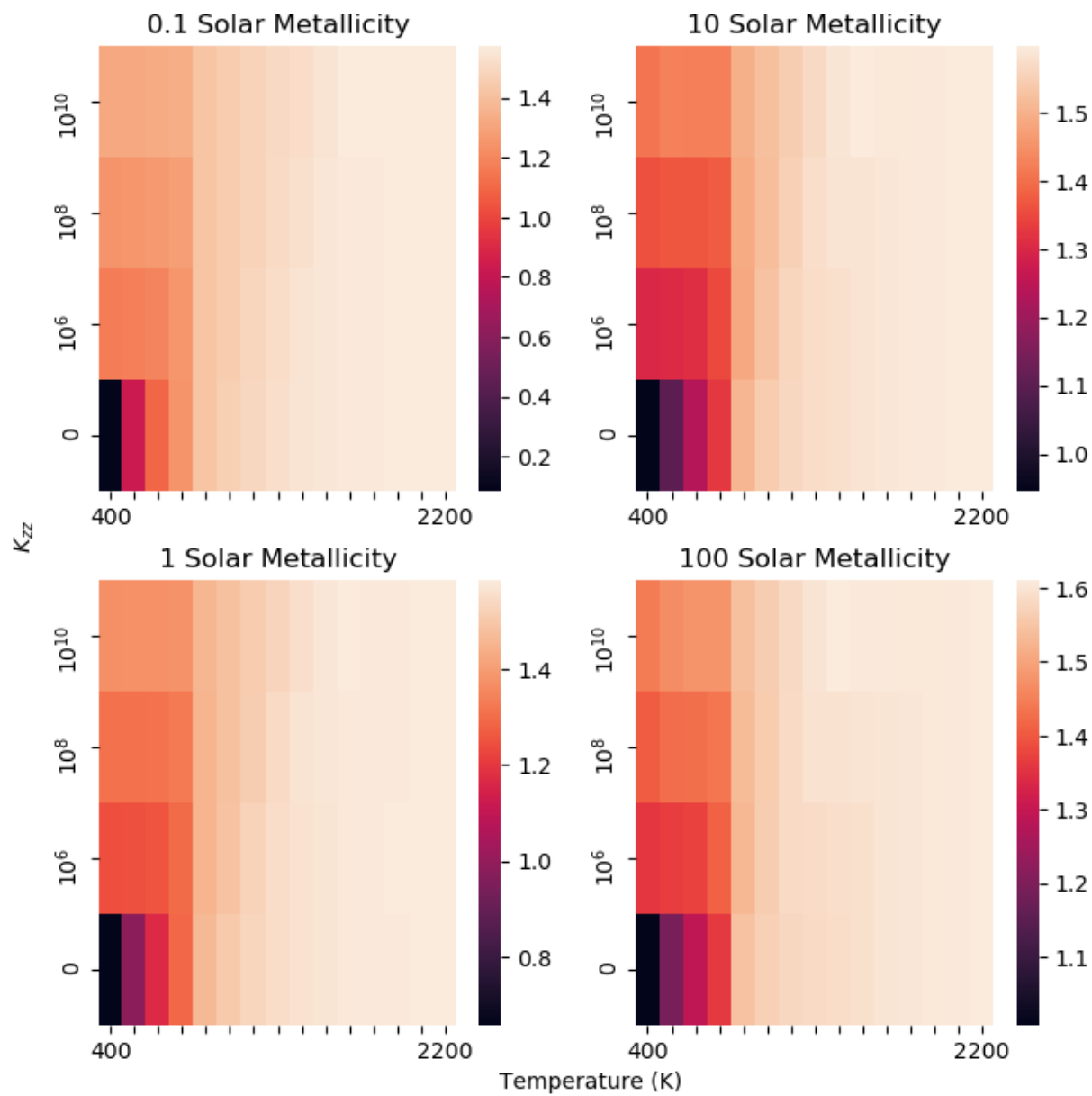

*Figure S9:* *Edge betweenness centrality as a function of temperature and $K_{zz}$, at 0.1, 1, 10, and 100 times solar metallicity. X-axis is temperature, y-axis is $K_{zz}$, and color corresponds to the weighted edge betweenness centrality.*

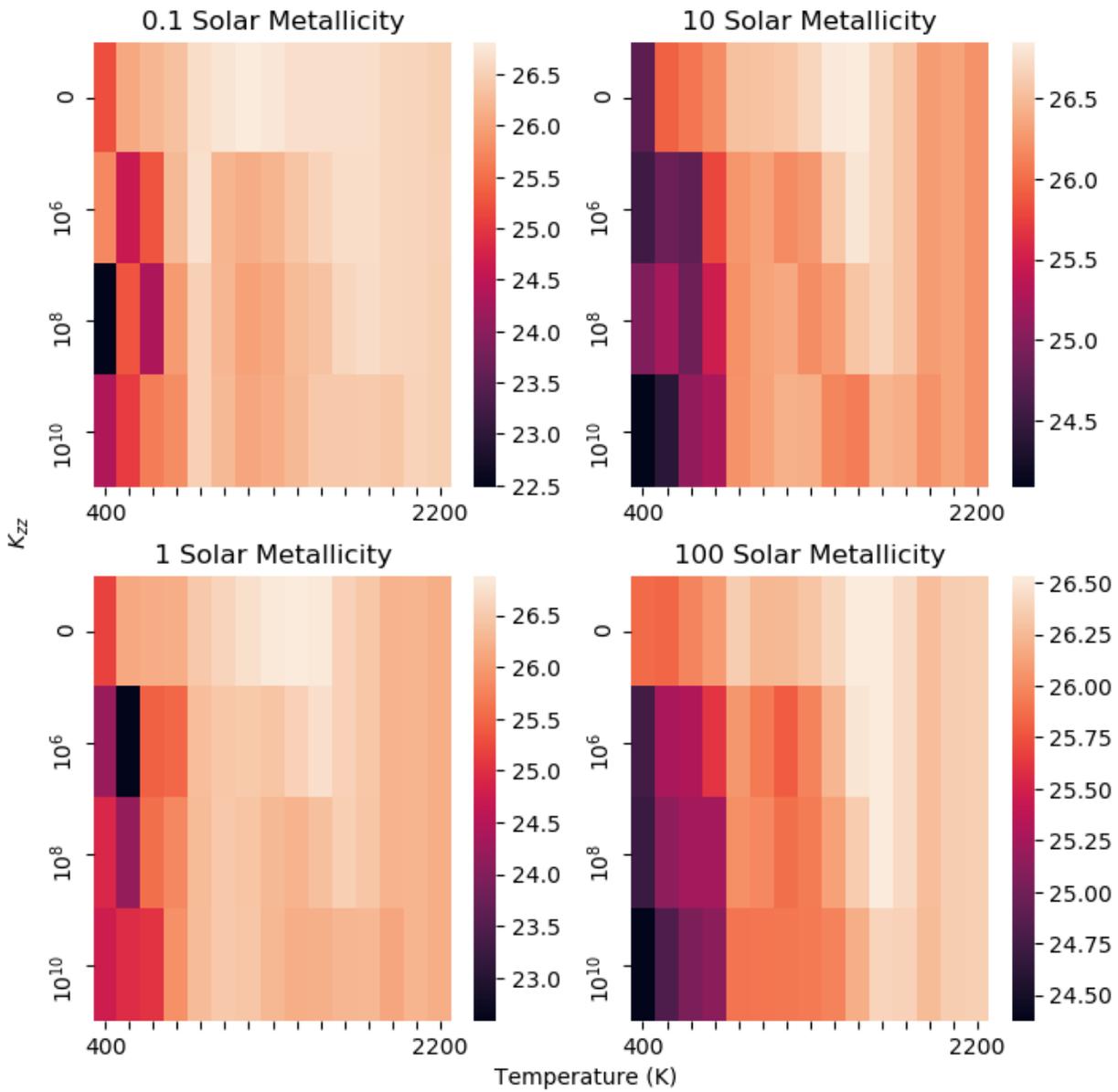

***Figure S10:*** *Average neighbor degree as a function of temperature and $K_{zz}$, at 0.1, 1, 10, and 100 times solar metallicity. X-axis is temperature, y-axis is $K_{zz}$, and color corresponds to the weighted average neighbor degree.*

## Comparison to Phi

The distribution of phi values also allowed distinguishing between different values of $K_{zz}$, though this was difficult to quantify due to the inability to use the KS test or Wasserstein distance metrics on the distributions.

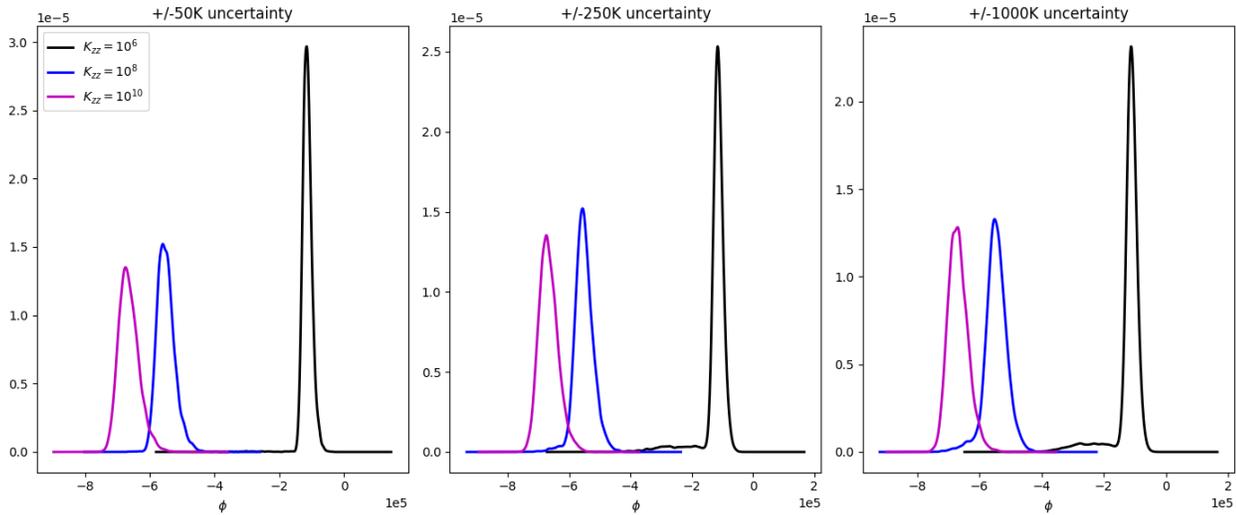

***Figure S11:*** *Phi distributions over a range of uncertainties and $K_{zz}$ values, drawn from a 10,000 point normal distribution of initial conditions centered at 1200K.*

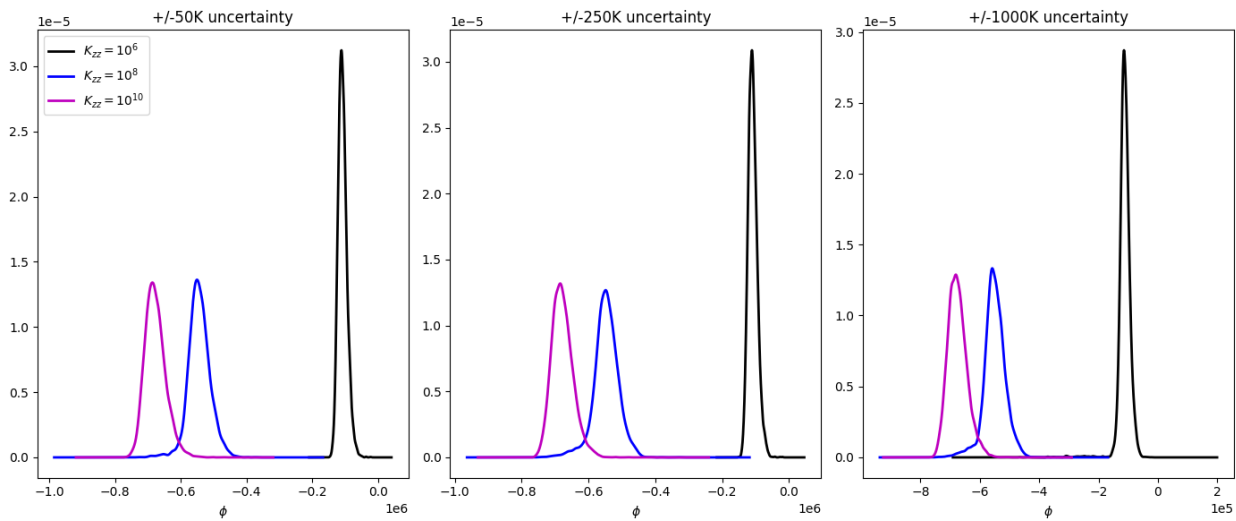

***Figure S12:*** *Phi distributions over a range of uncertainties and $K_{zz}$ values, drawn from a 10,000 point normal distribution of initial conditions centered at 2000K.*

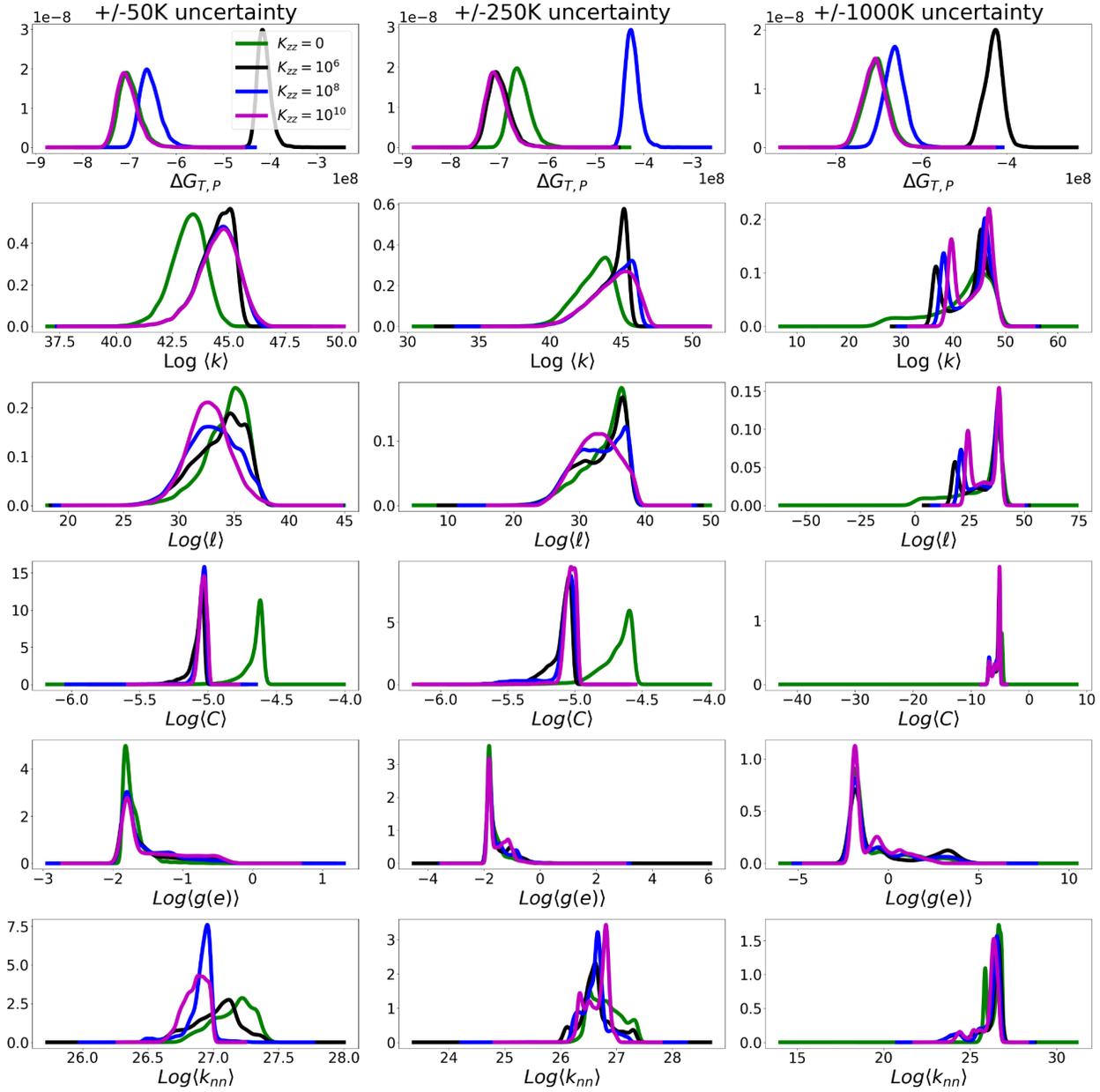

***Figure S13:*** *Distributions of network and thermodynamic measurements, interpolated from a 10,000 point normal distribution of initial conditions centered on a temperature of 1200K, at $K_{zz}$ values of 0, $10^6$, $10^8$, and $10^{10}$, and uncertainties in temperature of 50K, 250K, and 1000K. The bimodal distributions seen in the +/- 1000K distributions are an artifact of inserting noise into the system.*

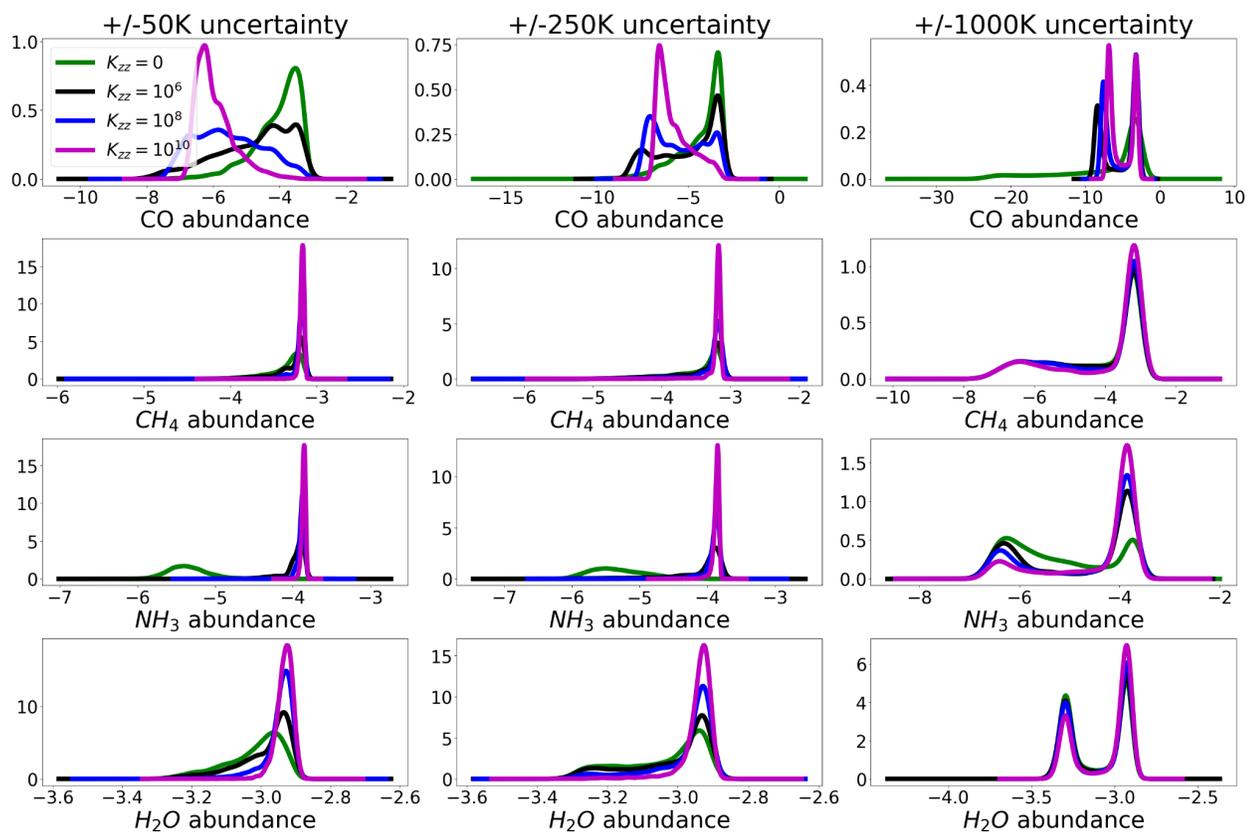

*Figure S14:* Distributions of molecular abundances, interpolated from a 10,000 point normal distribution of initial conditions centered on a temperature of 1200K, at $K_{zz}$ values of 0, $10^6$, $10^8$, and $10^{10}$, and uncertainties in temperature of 50K, 250K, and 1000K. The bimodal distributions seen in the +/- 1000K distributions are an artifact of inserting noise into the system.

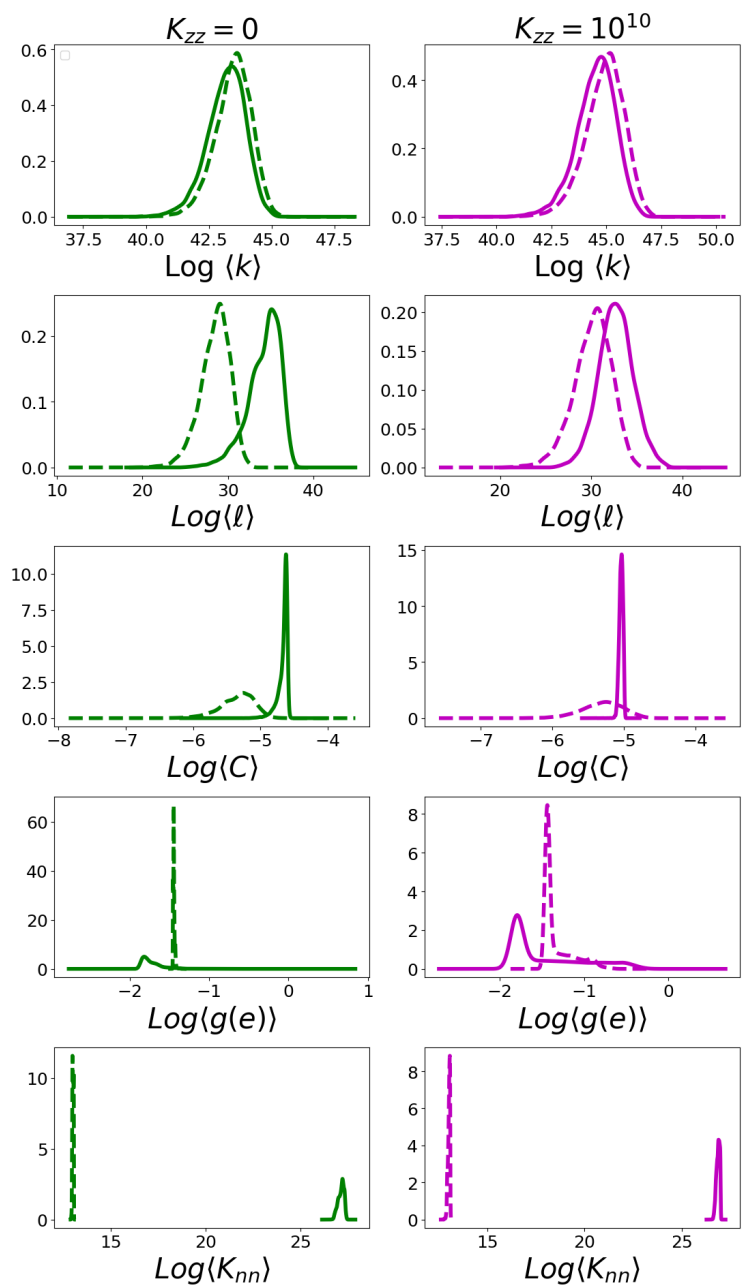

***Figure S15: T=1200K w/ Perturbations w/ all C-bearing compounds removed.*** *Distributions of network parameters when species containing carbon are removed from the network (dashed line) vs complete networks (solid line), at 1200K and 1 solar metallicity. The distributions of network parameters are easily distinguishable from each other, but the Gibbs free energy distribution is not. Interestingly, while the absolute values of the parameters change, the overall shape of them remains the same, despite the significant reduction in the network.*

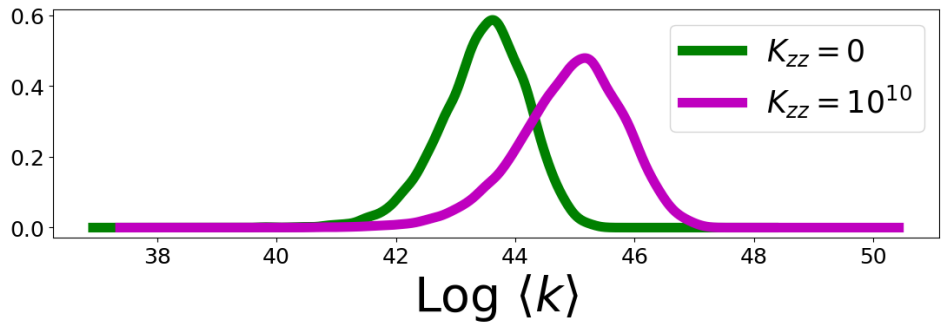
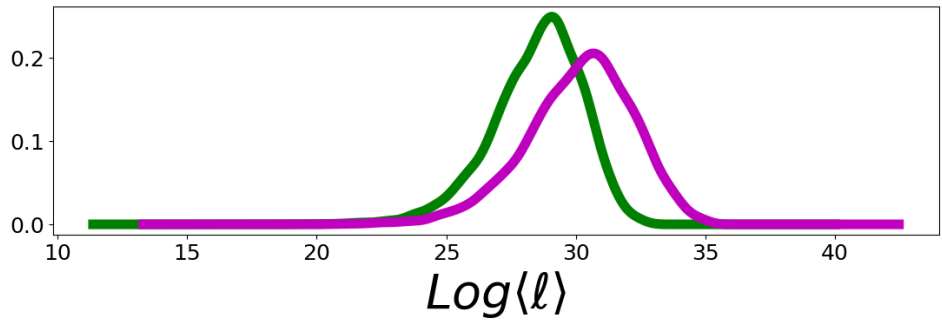
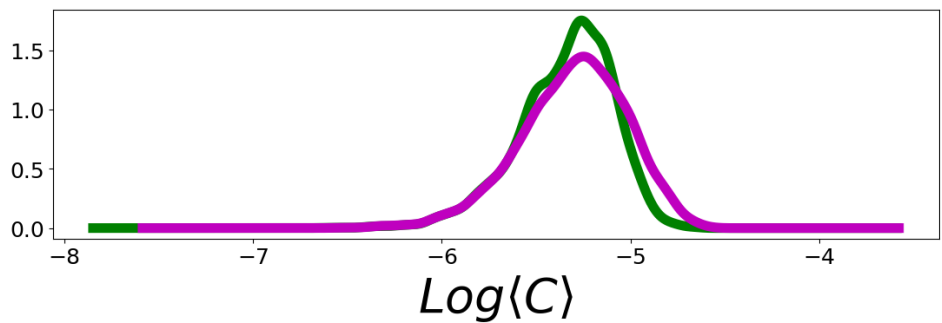
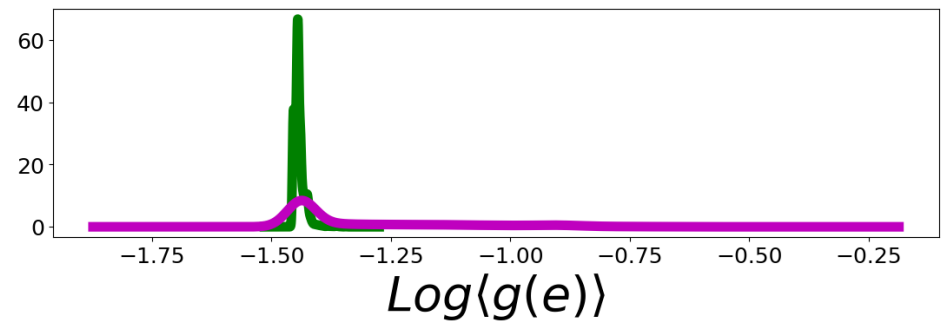
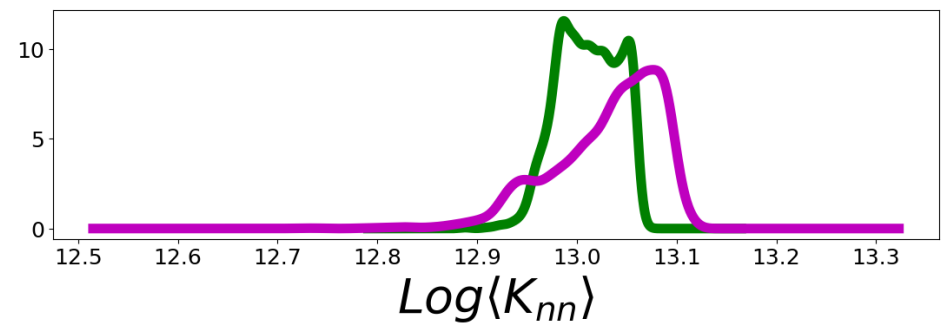

*Figure S16:* Distributions of network parameters when species containing carbon are removed from the network, at 1200K and 1 solar metallicity. While the absolute values of the parameters changed compared to the complete networks, the distributions are still distinguishable from each other as a function of $K_{zz}$.

Clustering Coefficient

| W.r.t $K_{zz}$ = | $K_{zz}$ = 0 | $K_{zz}$ = $10^6$ | $K_{zz}$ = $10^8$ | $K_{zz}$ = $10^{10}$ |
|---|---|---|---|---|
| $K_{zz}$ = 0 | 0 | | | |
| $K_{zz}$ = $10^6$ | 0.4059, $p$ = 0.0 | 0 | | |
| $K_{zz}$ = $10^8$ | 0.4049, $p$ = 0.0 | 0.0691, $p$ = 0.0 | 0 | |
| $K_{zz}$ = $10^{10}$ | 0.4014, $p$ = 0.0 | 0.1542, $p$ = 0.0 | 0.0952, $p$ = 0.0 | 0 |

*Table S1:* KS values for clustering coefficient for all interpolated distributions, with respect to different $K_{zz}$ values. Unsurprisingly, the equilibrium ($K_{zz}$ = 0) case is the most readily distinguishable from the other cases, but the varying levels of disequilibrium ($K_{zz}$ > 0) are still distinguishable from each other.

Mean Degree

| W.r.t $K_{zz}$ = | $K_{zz}$ = 0 | $K_{zz}$ = $10^6$ | $K_{zz}$ = $10^8$ | $K_{zz}$ = $10^{10}$ |
|---|---|---|---|---|
| $K_{zz}$ = 0 | 0 | | | |
| $K_{zz}$ = $10^6$ | 0.1638, $p$ = 0.0 | 0 | | |
| $K_{zz}$ = $10^8$ | 0.1951, $p$ = 0.0 | 0.1851, $p$ = 0.0 | 0 | |
| $K_{zz}$ = $10^{10}$ | 0.2455, $p$ = 0.0 | 0.2384, $p$ = 0.0 | 0.2366, $p$ = 0.0 | 0 |

*Table S2:* KS values for mean degree across all interpolated distributions, with respect to different $K_{zz}$ values. Distributions become more distinguishable from each other as the difference in $K_{zz}$ value between them increases.

Average Shortest Path Length

| W.r.t $K_{zz}$ = | $K_{zz}$ = 0 | $K_{zz}$ = $10^6$ | $K_{zz}$ = $10^8$ | $K_{zz}$ = $10^{10}$ |
|---|---|---|---|---|
| $K_{zz} = 0$ | 0 | | | |
| $K_{zz} = 10^6$ | 0.1725, $p = 0.0$ | 0 | | |
| $K_{zz} = 10^8$ | 0.2041, $p = 0.0$ | 0.1977, $p = 0.0$ | 0 | |
| $K_{zz} = 10^{10}$ | 0.2522, $p = 0.0$ | 0.2431, $p = 0.0$ | 0.2329, $p = 0.0$ | 0 |

***Table S3:*** *KS values for average shortest path length across all interpolated distributions, with respect to different $K_{zz}$ values. Distributions become more distinguishable from each other as the difference in $K_{zz}$ value between them increases.*

Average Neighbor Degree

| W.r.t $K_{zz}$ = | $K_{zz} = 0$ | $K_{zz} = 10^6$ | $K_{zz} = 10^8$ | $K_{zz} = 10^{10}$ |
|---|---|---|---|---|
| $K_{zz} = 0$ | 0 | | | |
| $K_{zz} = 10^6$ | 0.1695, $p = 0.0$ | 0 | | |
| $K_{zz} = 10^8$ | 0.1529, $p = 0.0$ | 0.0606, $p = 0.0$ | 0 | |
| $K_{zz} = 10^{10}$ | 0.1312, $p = 0.0$ | 0.0960, $p = 0.0$ | 0.0555, $p = 0.0$ | 0 |

***Table S4:*** *KS values for average shortest path length across all interpolated distributions, with respect to different $K_{zz}$ values. Average neighbor degree appears to be the weakest measurement, in terms of statistical distinguishability.*

Clustering coefficient,

| µ of distribution | $K_{zz} = 0$ | $K_{zz} = 10^6$ | $K_{zz} = 10^8$ | $K_{zz} = 10^{10}$ |
|---|---|---|---|---|
| 1200K, 50$_{sol}$ | 0 | 0.0038 | 0.0075 | 0.0051 |
| 2000K, 50$_{sol}$ | 0 | 0.0040 | 0.0092 | 0.01 |

Node Betweenness Centrality

| | | | | |
|---|---|---|---|---|
| 1200K, 50$_{sol}$ | 0 | 0.0013 | 0.0209 | 0.0661 |
| 2000K, 50$_{sol}$ | 0 | 0.0016 | 0.0551 | 0.0445 |

| Average Neighbor Degree | | | | |
|---|---|---|---|---|
| 1200K, 50$_{sol}$ | 0 | 0.0066 | 0.0063 | 0.0160 |
| 2000K, 50$_{sol}$ | 0 | 0.0081 | 0.0030 | 0.0625 |

*Table S5:* *The Wasserstein, or earth-mover distance, between the distributions of network measurement values, as a function of vertical mixing coefficient. This measurement provides one singular way of quantifying the statistical distinguishability between the distributions.*

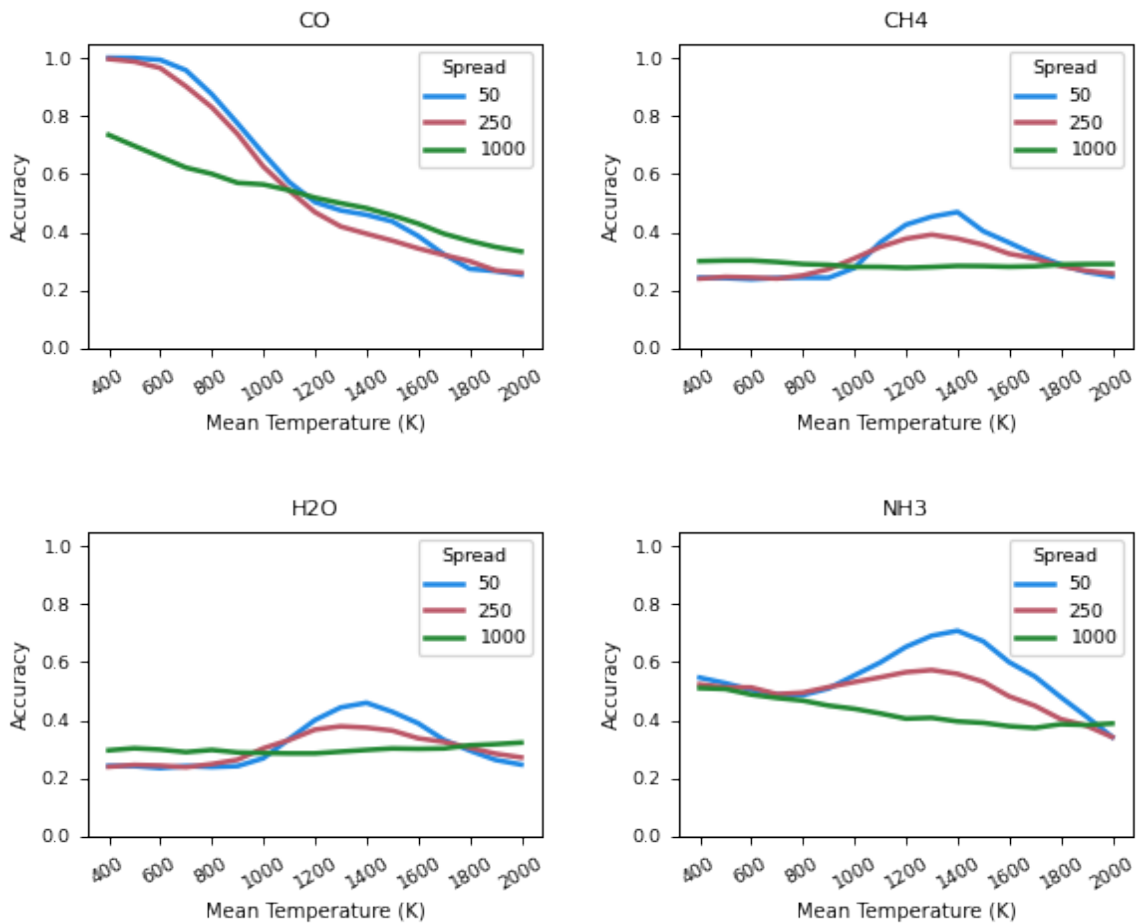

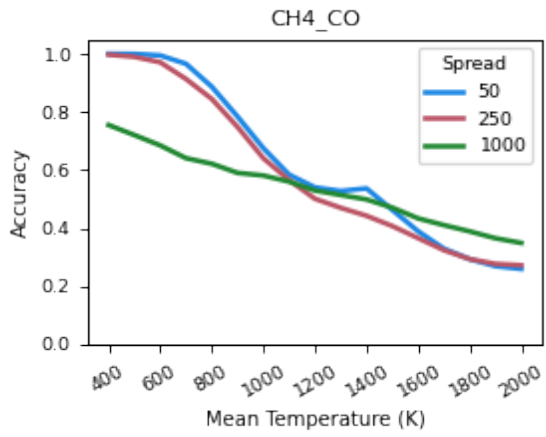
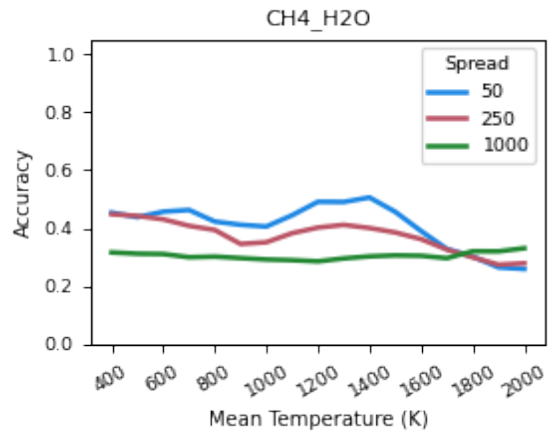
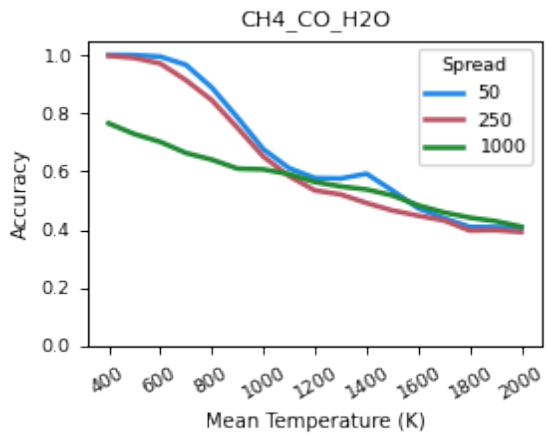
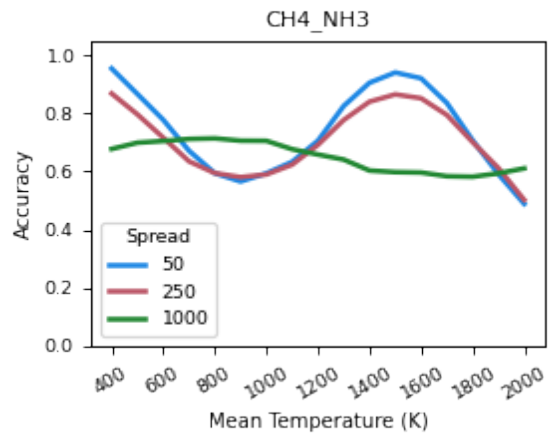
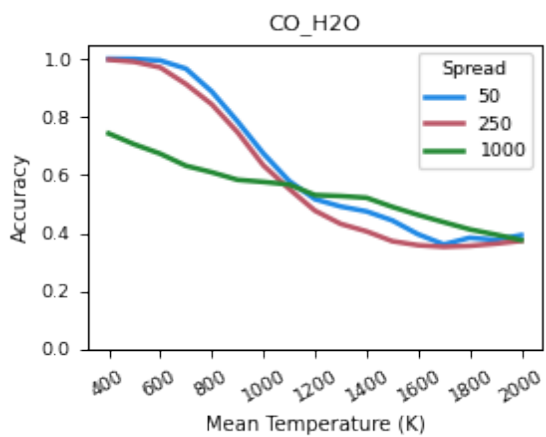
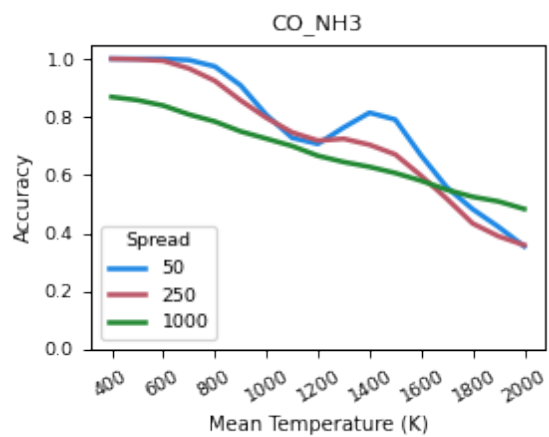

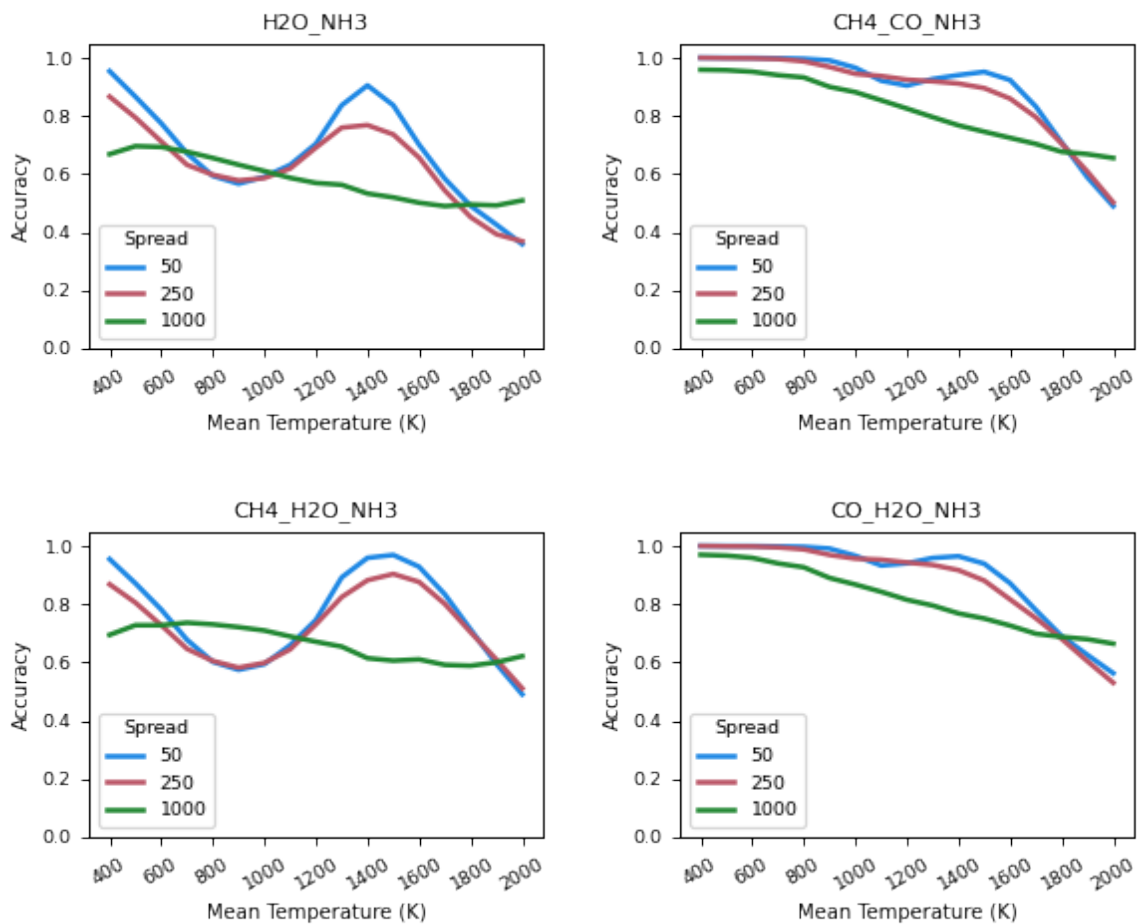

***Figure S17:** Accuracy of machine learning to predict $K_{zz}$ using the abundance of four species, alone and in various combinations. Predictions based on the abundance of one to two species fared poorly compared to topological, thermodynamic, and wider abundance measures. If three species were present, accuracy was raised, though still lower overall than that of the wider abundance measurements.*